\documentclass[aps,prd,twocolumn,superscriptaddress,showpacs]{revtex4-1}

\usepackage{lineno}
\usepackage{graphicx}

\begin{document}



\title{Production cross sections of hyperons and charmed baryons from $e^+e^-$
annihilation near $\sqrt{s} = 10.52$~GeV }


\noaffiliation

\affiliation{University of the Basque Country UPV/EHU, 48080 Bilbao}
\affiliation{Beihang University, Beijing 100191}
\affiliation{Budker Institute of Nuclear Physics SB RAS, Novosibirsk 630090}
\affiliation{Faculty of Mathematics and Physics, Charles University, 121 16 Prague}
\affiliation{Chonnam National University, Kwangju 660-701}
\affiliation{University of Cincinnati, Cincinnati, Ohio 45221}
\affiliation{Deutsches Elektronen--Synchrotron, 22607 Hamburg}
\affiliation{University of Florida, Gainesville, Florida 32611}
\affiliation{Department of Physics, Fu Jen Catholic University, Taipei 24205}
\affiliation{Justus-Liebig-Universit\"at Gie\ss{}en, 35392 Gie\ss{}en}
\affiliation{Gifu University, Gifu 501-1193}
\affiliation{SOKENDAI (The Graduate University for Advanced Studies), Hayama 240-0193}
\affiliation{Hanyang University, Seoul 133-791}
\affiliation{University of Hawaii, Honolulu, Hawaii 96822}
\affiliation{High Energy Accelerator Research Organization (KEK), Tsukuba 305-0801}
\affiliation{J-PARC Branch, KEK Theory Center, High Energy Accelerator Research Organization (KEK), Tsukuba 305-0801}
\affiliation{IKERBASQUE, Basque Foundation for Science, 48013 Bilbao}
\affiliation{Indian Institute of Science Education and Research Mohali, SAS Nagar, 140306}
\affiliation{Indian Institute of Technology Bhubaneswar, Satya Nagar 751007}
\affiliation{Indian Institute of Technology Guwahati, Assam 781039}
\affiliation{Indian Institute of Technology Madras, Chennai 600036}
\affiliation{Indiana University, Bloomington, Indiana 47408}
\affiliation{Institute of High Energy Physics, Chinese Academy of Sciences, Beijing 100049}
\affiliation{Institute of High Energy Physics, Vienna 1050}
\affiliation{Institute for High Energy Physics, Protvino 142281}
\affiliation{INFN - Sezione di Napoli, 80126 Napoli}
\affiliation{INFN - Sezione di Torino, 10125 Torino}
\affiliation{Advanced Science Research Center, Japan Atomic Energy Agency, Naka 319-1195}
\affiliation{J. Stefan Institute, 1000 Ljubljana}
\affiliation{Kanagawa University, Yokohama 221-8686}
\affiliation{Institut f\"ur Experimentelle Kernphysik, Karlsruher Institut f\"ur Technologie, 76131 Karlsruhe}
\affiliation{Kennesaw State University, Kennesaw, Georgia 30144}
\affiliation{King Abdulaziz City for Science and Technology, Riyadh 11442}
\affiliation{Department of Physics, Faculty of Science, King Abdulaziz University, Jeddah 21589}
\affiliation{Korea Institute of Science and Technology Information, Daejeon 305-806}
\affiliation{Korea University, Seoul 136-713}
\affiliation{Kyoto University, Kyoto 606-8502}
\affiliation{Kyungpook National University, Daegu 702-701}
\affiliation{\'Ecole Polytechnique F\'ed\'erale de Lausanne (EPFL), Lausanne 1015}
\affiliation{P.N. Lebedev Physical Institute of the Russian Academy of Sciences, Moscow 119991}
\affiliation{Faculty of Mathematics and Physics, University of Ljubljana, 1000 Ljubljana}
\affiliation{Ludwig Maximilians University, 80539 Munich}
\affiliation{University of Maribor, 2000 Maribor}
\affiliation{Max-Planck-Institut f\"ur Physik, 80805 M\"unchen}
\affiliation{School of Physics, University of Melbourne, Victoria 3010}
\affiliation{University of Miyazaki, Miyazaki 889-2192}
\affiliation{Moscow Physical Engineering Institute, Moscow 115409}
\affiliation{Moscow Institute of Physics and Technology, Moscow Region 141700}
\affiliation{Graduate School of Science, Nagoya University, Nagoya 464-8602}
\affiliation{Kobayashi-Maskawa Institute, Nagoya University, Nagoya 464-8602}
\affiliation{Nara Women's University, Nara 630-8506}
\affiliation{National Central University, Chung-li 32054}
\affiliation{National United University, Miao Li 36003}
\affiliation{Department of Physics, National Taiwan University, Taipei 10617}
\affiliation{H. Niewodniczanski Institute of Nuclear Physics, Krakow 31-342}
\affiliation{Nippon Dental University, Niigata 951-8580}
\affiliation{Niigata University, Niigata 950-2181}
\affiliation{Novosibirsk State University, Novosibirsk 630090}
\affiliation{Osaka City University, Osaka 558-8585}
\affiliation{Pacific Northwest National Laboratory, Richland, Washington 99352}
\affiliation{University of Pittsburgh, Pittsburgh, Pennsylvania 15260}
\affiliation{Research Center for Nuclear Physics, Osaka University, Osaka 567-0047}
\affiliation{Theoretical Research Division, Nishina Center, RIKEN, Saitama 351-0198}
\affiliation{RIKEN BNL Research Center, Upton, New York 11973}
\affiliation{University of Science and Technology of China, Hefei 230026}
\affiliation{Showa Pharmaceutical University, Tokyo 194-8543}
\affiliation{Soongsil University, Seoul 156-743}
\affiliation{Stefan Meyer Institute for Subatomic Physics, Vienna 1090}
\affiliation{Sungkyunkwan University, Suwon 440-746}
\affiliation{School of Physics, University of Sydney, New South Wales 2006}
\affiliation{Department of Physics, Faculty of Science, University of Tabuk, Tabuk 71451}
\affiliation{Tata Institute of Fundamental Research, Mumbai 400005}
\affiliation{Toho University, Funabashi 274-8510}
\affiliation{Department of Physics, Tohoku University, Sendai 980-8578}
\affiliation{Earthquake Research Institute, University of Tokyo, Tokyo 113-0032}
\affiliation{Department of Physics, University of Tokyo, Tokyo 113-0033}
\affiliation{Tokyo Institute of Technology, Tokyo 152-8550}
\affiliation{Tokyo Metropolitan University, Tokyo 192-0397}
\affiliation{Virginia Polytechnic Institute and State University, Blacksburg, Virginia 24061}
\affiliation{Wayne State University, Detroit, Michigan 48202}
\affiliation{Yamagata University, Yamagata 990-8560}
\affiliation{Yonsei University, Seoul 120-749}

\author{M.~Niiyama}\affiliation{Kyoto University, Kyoto 606-8502} 
\author{M.~Sumihama}\affiliation{Gifu University, Gifu 501-1193} 
\author{T.~Nakano}\affiliation{Research Center for Nuclear Physics, Osaka University, Osaka 567-0047} 

  \author{I.~Adachi}\affiliation{High Energy Accelerator Research Organization (KEK), Tsukuba 305-0801}\affiliation{SOKENDAI (The Graduate University for Advanced Studies), Hayama 240-0193} 
  \author{H.~Aihara}\affiliation{Department of Physics, University of Tokyo, Tokyo 113-0033} 
  \author{S.~Al~Said}\affiliation{Department of Physics, Faculty of Science, University of Tabuk, Tabuk 71451}\affiliation{Department of Physics, Faculty of Science, King Abdulaziz University, Jeddah 21589} 
  \author{D.~M.~Asner}\affiliation{Pacific Northwest National Laboratory, Richland, Washington 99352} 
  \author{V.~Aulchenko}\affiliation{Budker Institute of Nuclear Physics SB RAS, Novosibirsk 630090}\affiliation{Novosibirsk State University, Novosibirsk 630090} 
  \author{T.~Aushev}\affiliation{Moscow Institute of Physics and Technology, Moscow Region 141700} 
  \author{R.~Ayad}\affiliation{Department of Physics, Faculty of Science, University of Tabuk, Tabuk 71451} 
% ...
\collaboration{The Belle Collaboration}
\author{}



\begin{abstract}
We measure the inclusive production cross sections of hyperons and charmed baryons 
from $e^+e^-$ 
annihilation using a 800~fb$^{-1}$ data sample taken 
near the $\Upsilon(4S)$ resonance with the Belle detector
at the KEKB asymmetric-energy $e^+ e^-$ collider.
The feed-down contributions from heavy particles are subtracted using
our data, and the direct production cross sections are presented for the first time.
The production cross sections divided by the number of spin states for $S=-1$ hyperons 
follow an exponential function with a single slope parameter except for 
the $\Sigma(1385)^+$ resonance.
Suppression for $\Sigma(1385)^+$ and $\Xi(1530)^0$ hyperons is observed.
Among the production cross sections of charmed baryons,
a factor of three difference for $\Lambda_c^+$ states over $\Sigma_c$ states is observed.
This observation suggests a diquark structure for these baryons.

\end{abstract}


\maketitle

\section{Introduction}
\label{sec:intro}

Inclusive hadron production from $e^+e^-$ annihilation has been measured
for center-of-mass (CM) energy $\sqrt{s}$ of up to about 200~GeV, 
and summarized by the 
Particle Data Group~\cite{PDG2016}.
In $e^+e^-$ annihilation, hadrons are produced after the
$e^+e^-\rightarrow\gamma^*\rightarrow q\bar{q}$ creation
in the fragmentation process.
The observed production cross sections ($\sigma$) show an 
interesting dependence 
on their masses ($m$) and their angular momentum ($J$): 
$\sigma/(2J + 1) \propto \exp(-\alpha m)$,
 where $\alpha$ is a slope parameter.
The relativistic string fragmentation model~\cite{Andersson_PhysRept} 
reproduces well the angular and
momentum distributions of mesons in the 
fragmentation~\cite{Andersson_PhysRept}.
In this model, gluonic strings expand between the initial $q\bar{q}$ pair and
many $q\bar{q}$ pairs are created subsequently 
when the energy in the color field gets too large.
These $q\bar{q}$ pairs pick up other $\bar{q}q$ and form mesons
in the fragmentation process.


For the baryon production, two models are proposed: the diquark 
model~\cite{Andersson_PhysRept} and the popcorn
model~\cite{Andersson_PhysicaScripta}.
In the former, diquark ($qq$) and
anti-diquark ($\bar{q}\bar{q}$) pairs are created instead of a 
$q\bar{q}$.
In this model, a quark-quark pair is treated as 
an effective degree of freedom~\cite{Anselmino:1993}. 
{In the latter, three uncorrelated quarks are produced by either
$q\bar{q}$  creation
or diquark pair creation and then form baryons.}
{In Ref.}~\cite{Andersson_PhysicaScripta}, {the prediction of the
production rates by these models were compared, and they found
that, for decuplet baryons ($\Delta$ and $\Sigma(1385)$), 
the prediction by the diquark model is smaller than that by the 
popcorn model.
The production rates measured by ARGUS was compared with these 
models}~\cite{ARGUS_HYPERON_PLB1987},
{however, due to the large feed-down from heavier resonances, 
the direct comparison between the experimental data and the model
prediction was difficult.}

In earlier measurements at $\sqrt{s}=10$~GeV and at 
$\sqrt{s}=90$~GeV, production rates of most non-strange {light}
baryons and hyperons
follow an exponential mass dependence with a common slope parameter, 
but significant enhancements for $\Lambda$
and $\Lambda(1520)$ baryons are observed~\cite{PDG2016,Jaffe_Exotica}.
These enhancements could be explained by the light mass of the 
spin-0 diquark in $\Lambda$ baryons~\cite{Jaffe_Exotica,Wilczek_diquark}.  
However, the previous measurements of inclusive production cross sections contain
feed-down from heavier resonances. 
In order to compare the direct production cross sections of each baryon, 
feed-down contributions should be subtracted.
Charmed baryons have an additional interest, from the viewpoint of baryon 
structure: the color-magnetic interactions between the charm quark and 
the light quarks are suppressed due to the heavy charm quark mass, so that 
diquark degrees of freedom may be enhanced in the production mechanisms.

In this article, we report the production cross sections of hyperons and
charmed baryons using Belle~\cite{Belle_PTEP} data recorded at the KEKB 
$e^+e^-$ asymmetric-energy
collider~\cite{KEKB}. This high-statistics data sample has
good particle identification power.
{In this article, the direct cross sections of hyperons and charmed
baryons are described.}

This paper is organized as follows. In Sec.~\ref{sec:Analysis}, 
the data samples and the Belle detector are described, and the analysis 
to obtain the production cross sections is presented.
In Sec.~\ref{sec:Results}, the production cross sections are 
extracted for each baryon, and the production mechanism and the internal
structure of baryons are discussed. Finally, we summarize our results in 
Sec.~\ref{sec:Summary}.

\section{Analysis}
\label{sec:Analysis}

For the study of hyperon production cross sections in the hadronic events from $e^+e^-$ annihilation, 
we avoid contamination from $\Upsilon(4S)$ decay
by using off-resonance data taken at $\sqrt{s} = 10.52$~GeV, which is
60~MeV below the mass of the $\Upsilon(4S)$.
In contrast, for charmed baryons, for which the production
rates are small, especially for the excited states, 
we use both off- and on-resonance data, the latter recorded 
at the $\Upsilon(4S)$ energy ($\sqrt{s}=10.58$~GeV).
{In this article, we report the production cross sections of
$\Lambda$, $\Sigma^0$, $\Sigma(1385)^+$, $\Lambda(1520)$, $\Xi^-$, $\Omega^-$,
$\Xi(1530)^0$, $\Lambda_c^+$, $\Lambda_c(2595)^+$, $\Lambda_c(2625)^+$, 
$\Sigma_c(2455)^0$, $\Sigma_c(2520)^0$, $\Omega_c^0$, and $\Xi_c^0$.
These particles are reconstructed from charged tracks except for $\Sigma^0$.
Other ground-state baryons 
are omitted because their main decay modes contain neutral pions or neutrons.}
Since the absolute branching fractions for $\Omega_c^0$ and $\Xi_c^0$
are unknown, the production cross sections multiplied with the branching fractions
are presented.

The Belle detector is a large-solid-angle magnetic spectrometer that
consists of a silicon vertex detector (SVD), a central drift chamber (CDC), 
an array of aerogel threshold Cherenkov counters (ACC), 
 time-of-flight scintillation counters (TOF),
and an electromagnetic calorimeter (ECL) composed of CsI(Tl) crystals located
inside a superconducting solenoid coil that provides a 1.5~T magnetic field.
The muon/$K^0_L$ subsystem sandwiched within the solenoid's flux return is
not used in this analysis. 
The detector is described in detail elsewhere~\cite{AbashianNPA479,BrodzickaPTEP04D001}.


This analysis uses the data sets with two different inner detector
configurations.
A 2.0 cm beampipe and a three-layer silicon vertex detector (SVD1) were
used for the first samples of 140.0~fb$^{-1}$ (on-resonance) and 15.6~fb$^{-1}$
(off-resonance), while a 1.5 cm beampipe, a four-layer silicon detector (SVD2), 
and a small-cell inner drift chamber were used to record the remaining 571~fb$^{-1}$
(on-resonance) and 73.8~fb$^{-1}$ (off-resonance).

For the study of $S=-1$ hyperons, $\Lambda$, $\Sigma^0$, $\Sigma(1385)^+$
and $\Lambda(1520)$, which have relatively large production cross sections,
we use off-resonance data of the SVD2 configuration to avoid the 
systematic uncertainties due to the different experimental {setups}.
For the study of $S=-2$ and $-3$ hyperons, which have small cross sections,  
we use off-resonance data of the SVD1 and SVD2 configurations 
to reduce  statistical fluctuations.
For the study of charmed baryons, we use both off- and on-resonance data
taken with SVD1 and SVD2 configurations.
Since the charmed baryons from $B$-decay are forbidden in the high momentum
region due to the limited $Q$-value of 2.05~GeV for 
the 
$B^0\rightarrow {\bar{\Lambda}_c}^-p$ 
case and smaller for 
the excited states, 
we select prompt $c\bar{c}$ production events by selecting 
baryons with high momenta.


Charged particles produced from the $e^+e^-$ interaction point (IP) are selected
by requiring small impact parameters with respect to the IP
along the beam ($z$) direction and in the transverse plane
($r$--$\phi$) of $|dz| < 2\,{\rm cm}$ and  $dr<0.1$~cm, respectively.
For long-lived hyperons ($\Lambda$, $\Xi$, $\Omega$), we reconstruct their
trajectories and require consistency of the impact parameters to the IP as
described in the following subsections.
The particle identification is performed utilizing $dE/dx$ information 
from the CDC, time-of-flight measurements in the TOF, and 
Cherenkov light yield in the ACC.
The likelihood ratios for selecting $\pi$, $K$ and $p$ are 
required to be greater than 0.6 over the other particle hypotheses.
This selection has an efficiency of $90\sim 95$\% and a fake 
rate of $5\sim 9$\% ($\pi$ fakes $K$, for example).
Throughout this paper, {use of} charge-conjugate decay modes are implied, and
the cross sections of the sum of the baryon and anti-baryon production {is}
shown.
Monte Carlo (MC) events are generated using PYTHIA{6.2}~\cite{PYTHIA} and the
detector response is {simulated} using GEANT3~\cite{Geant3}.

We first obtain the inclusive differential cross sections ($d\sigma/dx_p$) 
as a function of hadron-scaled momenta, $x_p=pc/\sqrt{s/4-M^2c^4}$, where $p$ and $M$ are the
momentum and the mass, respectively,  of the particle. 
These distributions are shown after the
correction for the reconstruction efficiency and  {branching fractions}.
By integrating the differential cross sections in the $0\leq x_p \leq 1$ region, 
we obtain the cross section without radiative corrections (visible cross
sections). The QED radiative correction is applied in each $x_p$ bin of
the $d\sigma/dx_p$ distribution.
{The correction for the initial-state radiation (ISR) and the vacuum polarization 
is studied using PYTHIA by enabling or disabling these processes.
The final-state radiation (FSR) from charged hadrons is investigated using PHOTOS}~\cite{PHOTOS}.
The feed-down contributions from the heavier particles are subtracted from
the radiative-corrected total cross sections. Finally, the mass dependence of 
these feed-down-subtracted cross sections (direct cross sections) is investigated.

%
%
%
%
%

\subsection{$S=-1$ hyperons}

\begin{figure*}
\includegraphics[scale=0.7]{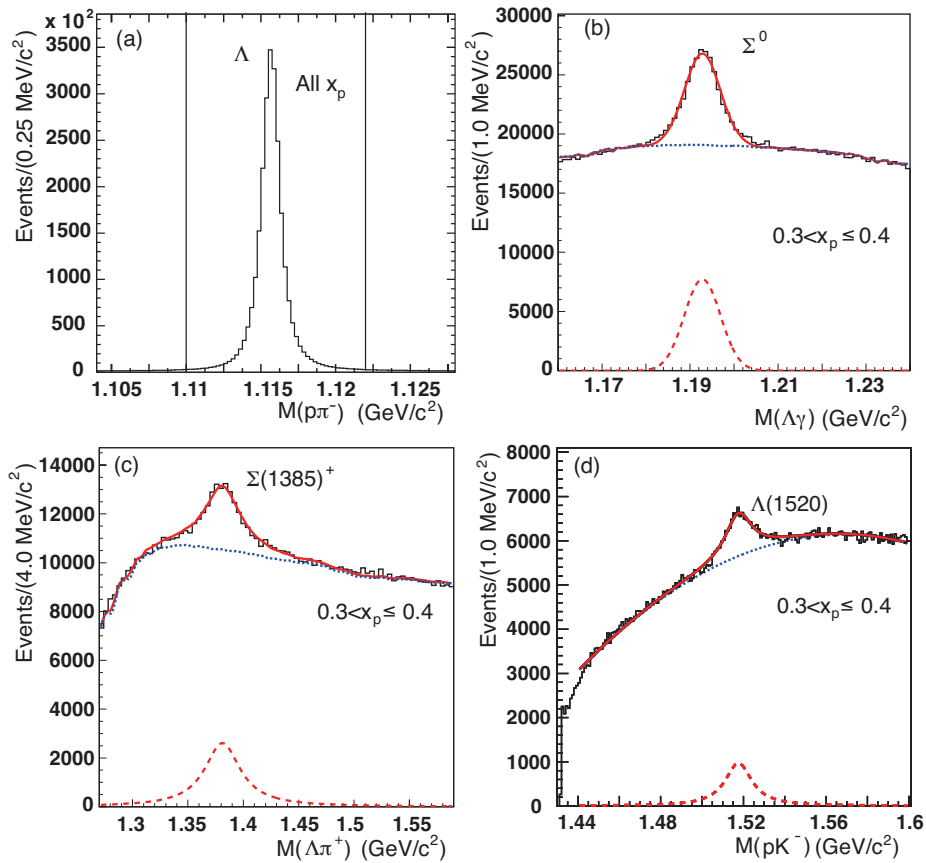}
\caption{ (a) The invariant-mass spectrum of ($p,\pi^-$). The vertical lines demarcate the signal
 region for $\Lambda$. (b), (c), (d) Invariant-mass spectra of
 ($\Lambda,\gamma$), ($\Lambda,\pi^+$), ($pK^-$), respectively.
 Fit results, signal shapes, and background shapes are shown by solid,
 dashed, and dotted curves, respectively.
}
\label{fig:invm_hyp1}
\end{figure*}

We start with the analysis of the $\Lambda$ baryon. 
We reconstruct a $\Lambda \rightarrow p \pi^-$ decay candidate from 
a proton and a pion candidate, and obtain the decay point and 
the momentum of the $\Lambda$.
The beam profile at the IP is wide in the horizontal direction ($x$) and
narrow in vertical ($y$);
the size of the IP region is typically $\sigma_x  \sim 100$~$\mu$m, 
$\sigma_y \sim 5$~$\mu$m, and $\sigma_z \sim 3$~mm~\cite{Belle_Vtx_resol}.
To select $\Lambda$ baryons that originate from the IP, we project 
the $\Lambda$ trajectory 
from its decay vertex toward the IP profile and then measure the
difference along the $x$ direction between its production point and the
IP centroid, 
$\Delta x$; we select events with $|\Delta x| < 0.2$~cm.
The $\Lambda$ candidates must have a flight length of 0.11~cm or more.
The invariant-mass spectrum of the surviving $p\pi^-$ combinations is shown in
Fig.~\ref{fig:invm_hyp1}(a).
We can see an almost background-free $\Lambda$ peak.
The events in the mass range of 1.110~GeV/$c^2$ $<M_\Lambda
<1.122$~GeV/$c^2$ are retained.
We investigate background events in
the sideband regions of
1.104~GeV/$c^2$$<M_\Lambda<1.110$~GeV/$c^2$ and 
1.122~GeV/$c^2$$<M_\Lambda<1.128$~GeV/$c^2$.
Due to the detector resolution, some signal events spill out of 
the mass range. 
This signal leakage is estimated using Monte Carlo (MC) 
 events, and is found to be about 4\% and 1\% of the events in the signal 
regions of $M_\Lambda$ and $\Delta x$, respectively.
The MC study also shows that background events distribute rather evenly
both in the invariant mass and $\Delta x$. Therefore, the
background contributions are estimated by the sum of sideband events
after subtracting the signal leakage.

Next, a $\Lambda$ candidate is combined with a photon or a $\pi^+$ to 
form 
a $\Sigma^0$ or a $\Sigma(1385)^+$ candidate, respectively. 
The energy of the photon from the $\Sigma^0$ decay must exceed 
 45~MeV {to suppress backgrounds.}
The invariant-mass spectra
of the $\Lambda \gamma$ and $\Lambda \pi^+$ combinations are shown in 
Figs.~\ref{fig:invm_hyp1}(b) and \ref{fig:invm_hyp1}(c), 
where peaks of $\Sigma^0$ and
$\Sigma(1385)^+$ are observed.
{Background shapes ($h(m)$) for $\Sigma^0$ and $\Sigma(1385)$
as functions of the invariant mass ($m$) are obtained using MC events of
$e^+e^- \rightarrow q\bar{q}$ production, where $q=u,d,s,c$.
We apply Wiener filter~}\cite{NumericalRevipes} {for $h(m)$ to avoid fluctuation 
due to the finite statistics of MC samples.
For the fit the MC spectra to the real data, 
we multiply the first order polynomial function ($a m+b$) to $h(m)$, 
where $a$ and $b$ are free parameters.}
The signal yields of $\Sigma^0$ are estimated by fitting the $\Lambda
\gamma$ spectrum in the range
{1.17~GeV/$c^2$$<M_{\Lambda\gamma}<$1.22~GeV/$c^2$} 
with a Gaussian and the background spectrum, where all parameters
are determined from the fit. 
In this analysis, all fit parameters are floated in each $x_p$ bin
unless otherwise specified.
Note that the mass resolution for
the signal is almost entirely determined by the energy resolution of 
the low-energy photon and can be approximated by a Gaussian shape.
On the other hand, a non-relativistic Breit-Wigner function is used to 
estimate the signal yields of $\Sigma(1385)^+$ since the detector resolution
is negligible compared to the natural width.
The fit region is {1.3~GeV/$c^2$$<M_{\Lambda\pi^+}<$1.5~GeV/$c^2$}, and all
parameters are floated in the fit.

For the reconstruction of $\Lambda(1520) \rightarrow K^- p$, 
tracks identified as a kaon and a proton, each with a small impact
parameter with respect to the IP, are selected.
The invariant-mass spectrum of $K^- p$ pairs is shown in 
Fig.~\ref{fig:invm_hyp1}(d). A clear peak of the $\Lambda(1520)$ is 
seen above the
combinatorial background.
We employ a third-order polynomial for the background and a non-relativistic
Breit-Wigner function to estimate the signal yields, where all
parameters are floated except for the width of the Breit-Wigner
function, which is fixed to the PDG value to stabilize the fit.
The fit region is {1.44~GeV/$c^2<M_{K^-p}<1.6$~GeV/$c^2$.}

\subsection{$S=-2,-3$ hyperons}

\begin{figure*}
\includegraphics[scale=0.7]{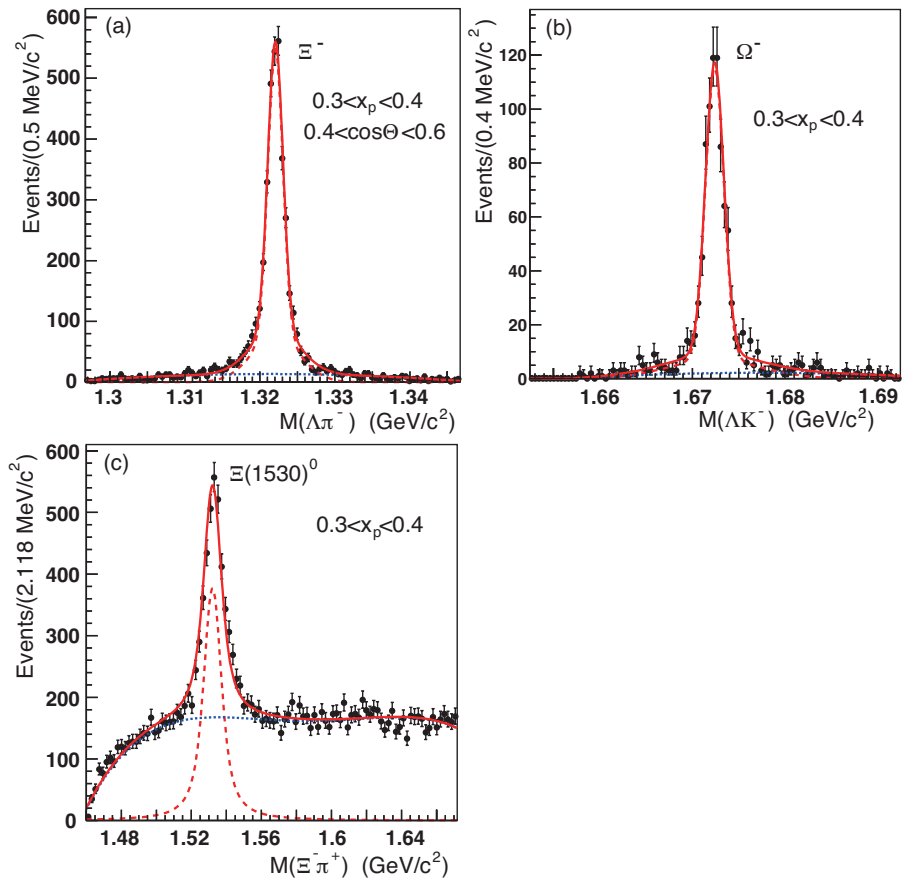}
\caption{ (a)-(c) Reconstructed mass spectra for $S=-2$ and $-3$
hyperon candidates.
{ Fit results, signal shapes, and background shapes are shown by solid,
 dashed, and dotted curves, respectively.}
}
\label{fig:invm_hyp2}
\end{figure*}

The $\Xi^-$ and $\Omega^-$ are reconstructed from $\Xi^- \rightarrow \Lambda
\pi^-$ and $\Omega^- \rightarrow \Lambda K^-$ decay modes, respectively.
We reconstruct the vertex point of a $\Lambda \rightarrow p \pi^-$ 
candidate, as before, but do not impose the IP constraint on $\Delta x$
here to account for the long lifetime of the  $S=-2,-3$ hyperons.
Instead, the trajectory of the $\Lambda$ is combined with a $\pi^-$ ($K^-$) 
and the helix trajectory of the
$\Xi^-$ ($\Omega^-$) candidate is reconstructed. 
This helix is extrapolated back toward the IP. The distance of
the generation point of the $\Xi^-$ ($\Omega^-$) from IP {along the
radial ($dr$) and the beam direction ($dz$)}
must satisfy $dr<0.1$ (0.07)~cm and $|dz|<2.0$ (1.1)~cm.
The invariant-mass spectra of $\Lambda \pi^-$ and $\Lambda K^-$ pairs
are shown in Figs.~\ref{fig:invm_hyp2}(a) and \ref{fig:invm_hyp2}(b). 
We see prominent peaks of $\Xi^-$ and $\Omega^-$.
The $\Xi(1530)^0$ hyperon candidates are reconstructed 
from $\Xi^-\pi^+$ pairs, whose invariant mass is shown in 
Fig.~\ref{fig:invm_hyp2}(c).

Signal peaks of $\Xi^-$ and $\Omega^-$ are fitted with double-Gaussian
functions, and those of $\Xi(1530)^0$ are fitted with Voigt functions.
A second-order Chebyshev polynomial is used to describe background contributions.
{All parameters are floated.}
The fit regions are 1.28~GeV/$c^2<M_{\Lambda\pi^-}<1.375$~GeV/$c^2$,
1.465~GeV/$c^2<M_{\Xi^-\pi^+}<1.672$~GeV/$c^2$, and
1.652~GeV/$c^2<M_{\Lambda K^-}<1.692$~GeV/$c^2$ for
$\Xi^-$, $\Xi(1530)^0$, and $\Omega^-$, respectively.
The widths of $\Xi(1530)^0$ obtained by the fit are consistent with
the PDG value.

\subsection{Charmed baryons}
\label{sec:ana_charmed_baryons}

%

\begin{figure*}
\includegraphics[scale=0.8]{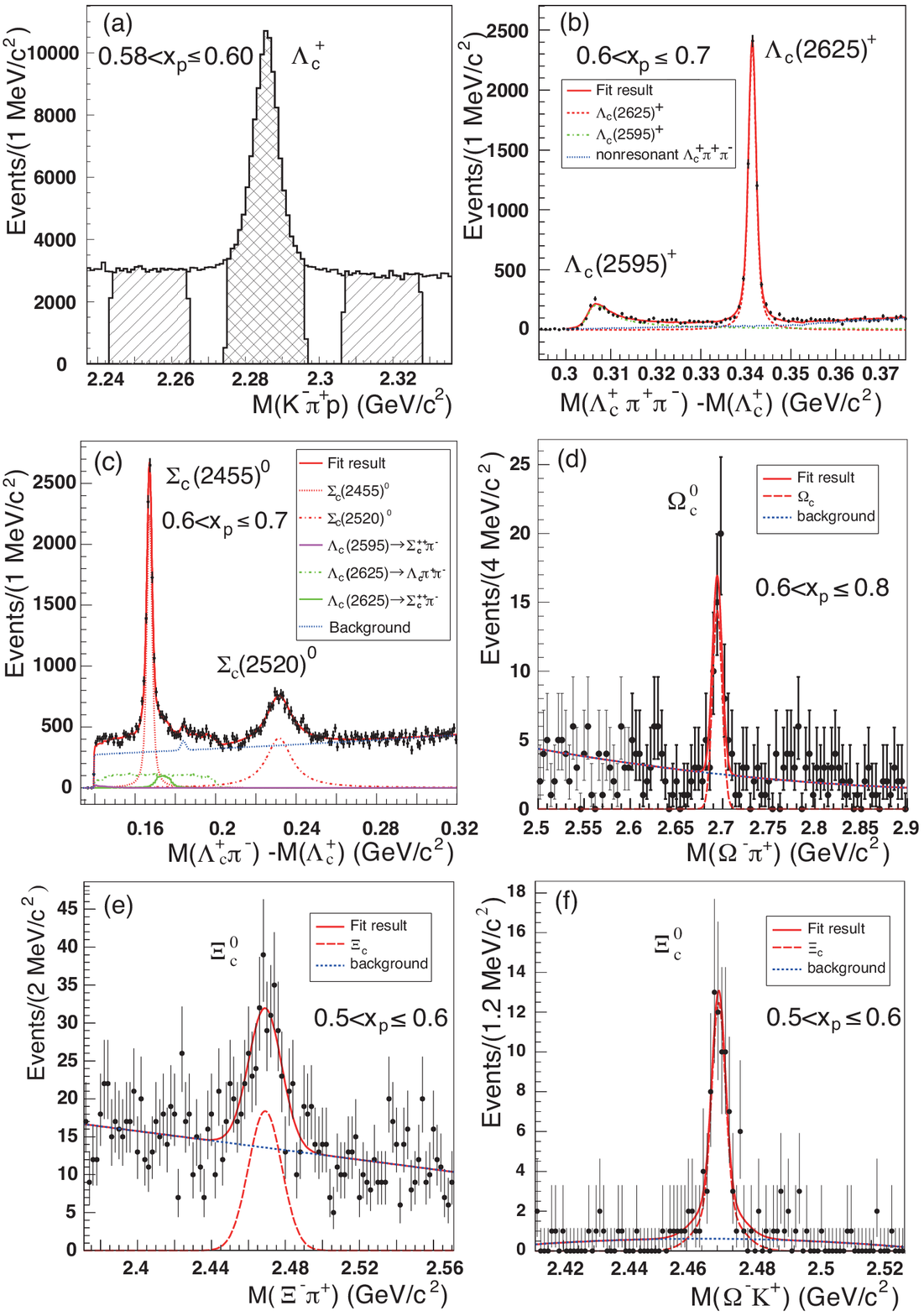}
\caption{Reconstructed mass spectra  of charmed baryon candidates.
(a) The invariant-mass spectrum of $\Lambda_c^+$. The signal and sideband 
regions are indicated by the double-hatched and hatched histograms, respectively.
(b) The mass-difference distribution of $\Lambda_c^+\pi^+\pi^- -\Lambda_c^+$.
(c) The mass-difference distribution for {$\Lambda_c^+\pi^-$-$\Lambda_c^+$}. 
(d) The invariant-mass spectrum of $\Omega^-\pi^+$.
(e), (f) The invariant-mass spectra of $\Xi^-\pi^+$ and $\Omega^-K^+$, respectively.
}
\label{fig:invm_c}
\end{figure*}

For the study of charmed baryons,
we use both off- and on-resonance data, the latter recorded
at the $\Upsilon(4S)$ energy ($\sqrt{s}=10.58$~GeV).
To eliminate the $B$-meson decay contribution, 
the charmed-baryon candidates are required to have
$x_p>0.44$ in the on-resonance data.
For the reconstruction of charmed baryons, we apply the same PID 
and impact parameter criteria as for hyperons.

\begin{figure}
\includegraphics[scale=0.4]{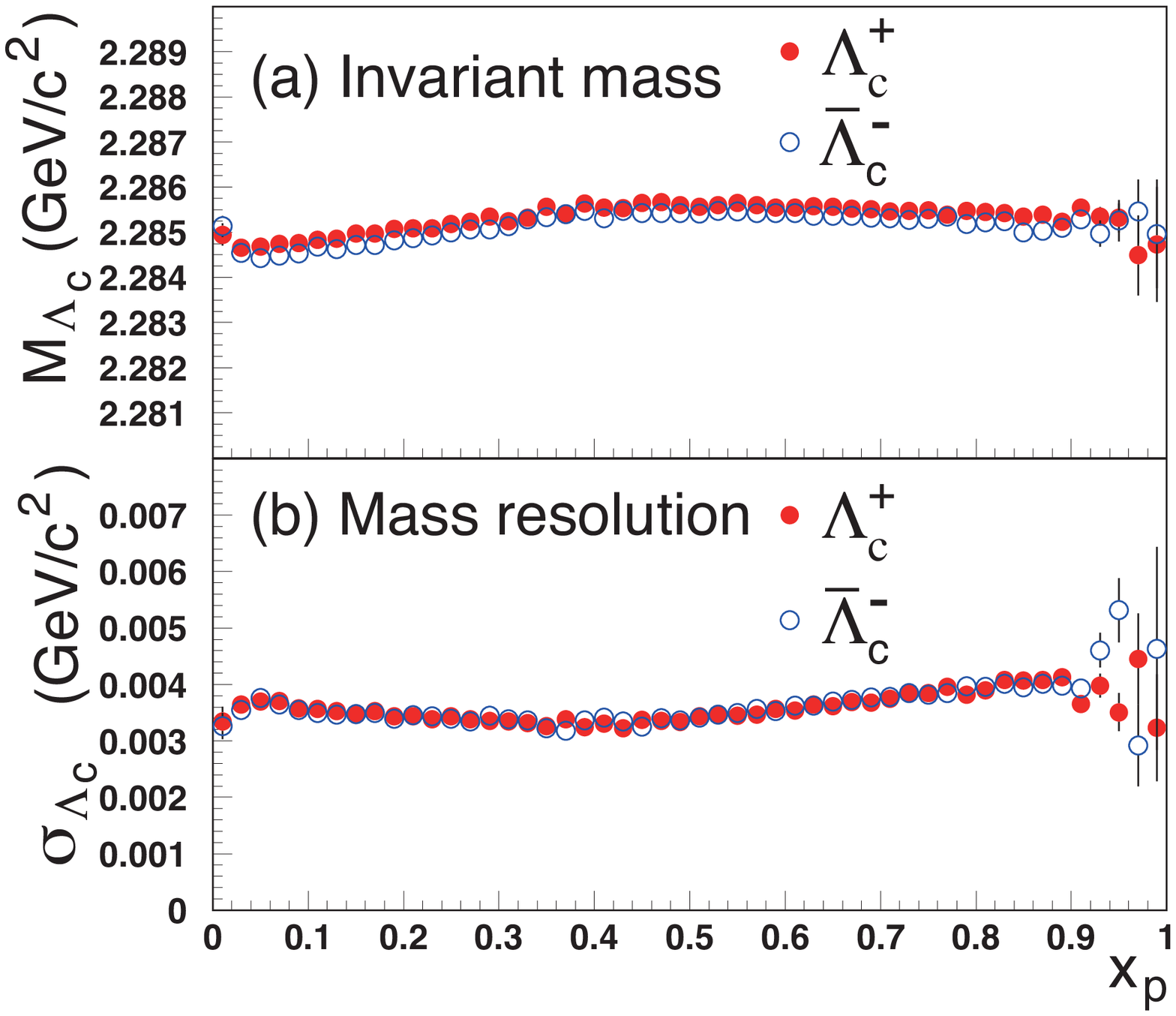}
\caption{Invariant mass (a) and R.M.S. width (b) of the 
fitted $\Lambda_c^+$ and $\bar{\Lambda}_c^-$ as a function of $x_p$.}
\label{fig:pl_mw_Lc}
\end{figure}

First, we reconstruct the $\Lambda_c^+$ baryon in 
the $\Lambda_c^+\rightarrow \pi^+K^- p$ decay mode.
To improve the momentum resolution, 
we apply a vertex-constrained fit that incorporates the IP profile.
We fit the invariant-mass spectra in 50 $x_p$ bins (Fig.~\ref{fig:invm_c}(a)), 
and obtain peak positions and widths of $\Lambda_c^+$ as a function of
the momentum (Fig.~\ref{fig:pl_mw_Lc}).
The peak positions are slightly smaller than the PDG value by 1$\sim$1.4~MeV/$c^2$.
In order to avoid misestimation of the yields, 
we select $\Lambda_c^+$ candidates whose mass ($M$) is within 
$3\sigma$ of the peak of a Gaussian fit ($M_{\Lambda_c}(x_p)$) 
as signal. Candidates with
$-11\sigma < |M-M_{\Lambda_c}(x_p)-3~{\rm MeV }/c^2|< -5\sigma$ and 
$+5\sigma < |M-M_{\Lambda_c}(x_p)+3~{\rm MeV }/c^2|< 11\sigma$
are treated as sideband.
We estimate background yields under the signal peak from the yields in 
the sidebands, and correct for reconstruction efficiency 
using MC $e^+e^- \rightarrow c\bar{c}$ events.
In the $\Lambda_c^+ \rightarrow \pi^+K^-p$ decay, 
the intermediate resonances ($K(890)^0$, $\Delta$, and $\Lambda(1520)$) can
contribute, and the distribution in the Dalitz plane is not uniform~\cite{Belle_Lc_double_cabbibo}.
To avoid the uncertainty in the reconstruction efficiency
correction due to these intermediate states, the correction is applied for the Dalitz
distribution of $\Lambda_c^+$ signal region after subtracting the
sideband events.
In the low $x_p$ region ($x_p\leq 0.44$), we obtain the cross section using
off-resonance data, whereas we utilize both off- and on- resonance data
in the high $x_p$ region ($x_p>0.44$).

We reconstruct $\Sigma_c^{(*)0}$ or excited $\Lambda_c^{*+}$ states
by combining a 
 $\Lambda_c^+$ candidate with a $\pi^-$ or a $\pi^+\pi^-$ pair, respectively. 
Among several $\Lambda_c^+\pi^-$ ($\Lambda_c^+\pi^+\pi^-$) combinations in
one event, we select the one with the best fit quality in the 
vertex-constraint fit.
The background events are subtracted using the sideband distribution, 
as described above. Reconstructed invariant-mass spectra of 
$\Delta M(\pi\pi)={M}(\Lambda_c^+\pi^+\pi^-) -{M}(\Lambda_c^+)$
and $\Delta M(\pi^-)={M}(\Lambda_c^+\pi^-) -
{M}(\Lambda_c^+)$ are shown in Figs.~\ref{fig:invm_c}(b) and 
\ref{fig:invm_c}(c), respectively.
We see clear peaks of
$\Lambda_c(2595)^+$ and $\Lambda_c(2625)^+$ in Fig.~\ref{fig:invm_c}(b) and
of $\Sigma_c(2455)^0$ and $\Sigma_c(2520)^0$ in Fig.~\ref{fig:invm_c}(c). 
Since the peaks of these states are not statistically significant in the 
low $x_p$ region of the 
off-resonance data, we obtain the cross section
in the $x_p>0.4$ region and extrapolate to the entire 
$x_p$ region using the Lund fragmentation model. 
The fragmentation-model dependence introduces a systematic uncertainty
that is estimated by the variation using other models.
The yields of these charmed baryons are obtained from fits to invariant-mass
distributions in the mass range 0.28~GeV/$c^2<\Delta M({\pi\pi})<0.38$~GeV/$c^2$
and 0.145~GeV/$c^2<\Delta M({\pi^-})<0.32$~GeV/$c^2$ for excited
$\Lambda_c$ baryons and $\Sigma_c$ baryons, respectively.

In the $\Delta M(\pi\pi)$ spectra, the background shape can be described by
the combination of $\Lambda_c^+$ with pions that are not associated
with resonances.
We generate inclusive $e^+ e^- \to \Lambda_c^+ X$ MC events, 
and use the invariant mass of $\Lambda_c^+\pi^+\pi^-$ combinations to 
describe the background spectra.
We use a Voigtian~\cite{Voigt} function 
to describe the line-shape of $\Lambda_c(2625)^+$, 
where the width and the 
resolution are set as free parameters.
The widths obtained by the fit are smaller than 1~MeV/$c^2$,
and are consistent with the upper limit ($0.97$~MeV/c$^2$) in the PDG.
The mass of $\Lambda_c(2595)^+$ is very close to the mass threshold of
$\Lambda_c^+\pi^+\pi^-$ and so the line shape is asymmetric.
We use the theoretical model of Cho~\cite{ChoPRD50} to 
describe the line-shape of $\Lambda_c(2595)^+$, with parameters
obtained by CDF~\cite{CDF_PRD84012003}.
This model describes the width of the $\Lambda_c(2595)^+$ as a function of
the mass, and produces a long tail in the high-mass region.
To reduce the systematic uncertainty due to the tail contribution, we 
evaluate the yield of the $\Lambda_c(2595)^+$ in the 
$\Delta M(\pi\pi)<0.33$~GeV/$c^2$ region.
The systematic uncertainty due to this selection
is estimated by changing the $\Delta M(\pi\pi)$ region, and is included in the
systematic due to the signal shape in Table~\ref{tbl:syst_charm}.

We also use Voigtian functions to describe $\Sigma_c(2455)^0$ 
and $\Sigma_c(2520)^0$; the Belle measurements~\cite{Belle_ScWidth} of 
the masses and widths are used.
The fit results are shown in Figs.~\ref{fig:invm_c}(b) and 
\ref{fig:invm_c}(c). 
In Fig.~\ref{fig:invm_c}(c), the background spectrum
exhibits a non-uniform structure due to the feed-down contribution from 
$\Lambda_c(2595)^+$ and $\Lambda_c(2625)^+$.
These resonances decay into $\Lambda_c^+\pi^+\pi^-$, $\Sigma_c^{++}\pi^-$
$\Sigma_c^{+}\pi^0$, and  $\Sigma_c^{0}\pi^+$, where
$\Lambda_c^+\pi^+\pi^-$ and $\Sigma_c^{++}\pi^-$ modes are considered
background. Feed-down contributions from $\Lambda_c^+$ excited states to
the $\Sigma_c^{0}\pi^+$ mode is subtracted later.
In the  $\Delta M(\pi^-)$ spectra of the
$\Lambda_c^* \rightarrow\Lambda_c^+\pi^+\pi^-$, $\Sigma_c^{++}\pi^-$
reactions from MC simulation, a small enhancement
around $\Delta M(\pi^-)=0.187$~GeV/$c^2$ is likely due to the contribution
from $\Xi_c^0$ as discussed in Ref.~\cite{Belle_ScWidth}, and a Gaussian
function is used to describe this peak.
The magnitude of background contributions is treated as
a free parameter, and a fit including the signal peaks
is shown in Fig.~\ref{fig:invm_c}(c).
The $\chi^2$ per the number of degrees of freedom ($ndf$) values are in the range 
from 148/163 to 203/163, and are
reasonably good in each $x_p$ bin; deviations
from the fit function are within statistical uncertainties.

Figure~\ref{fig:invm_c}(d) shows the invariant-mass spectrum of 
$\Omega^-\pi^+$ pairs, where a peak corresponding to  $\Omega_c^0$ is seen.
The yields of $\Omega_c^0$ are obtained from fits to invariant-mass
distributions in the range of 2.5~GeV/$c^2<M_{\Omega_c^0}<2.9$~GeV/$c^2$. 
The signal and background shapes are described by Gaussian functions and 
second-order Chebyshev polynomial functions, 
{where the mean and width of Gaussian functions are allowed to float.}
$\Xi_c^0$ baryons are reconstructed in two decay modes:
$\Xi_c^0\rightarrow \Xi^-\pi^+$ and $\Xi_c^0\rightarrow \Omega^-K^+$, as
shown in Figs.~\ref{fig:invm_c}(e) and \ref{fig:invm_c}(f).
The yields of $\Xi_c^0$ are obtained from fits to invariant-mass
distributions in the range of 2.321~GeV/$c^2<M_{\Xi_c^0}<2.621$~GeV/$c^2$. 
The signal and background shapes are described by double-Gaussian functions and 
second-order Chebyshev polynomial functions.

\subsection{Inclusive cross sections}

\begin{table*}[htb]
  \caption{Total cross sections before (visible) and after the radiative correction,
where the first and second errors are statistical and systematic uncertainties.
For the charmed strange baryons, the cross sections times branching fractions 
are listed.}
  \label{tbl:cross_inclusive_radCor}
\begin{tabular}{ l c c c c c c c c} \hline\hline
Particle &  Mode & Branching&  Visible     & Radiative corrected &  Ratio of \\
         &       & fraction &   cross      &   cross & before and after\\
         &       & (\%)     & section (pb) & section (pb) &the correction\\ \hline
$ \Lambda $ &  $p\pi^-$  &  $63.9\pm 0.5$    & $308.80\pm 0.37\pm 17$ &  $276.50\pm 0.33\pm 16$ &0.895 \\
$ \Lambda(1520) $ & $pK^-$  &  $22.5\pm 0.5$ & $14.32\pm0.23\pm 1.0$ &  $12.80\pm 0.20\pm 0.94$ &0.894 \\
$ \Sigma^0 $ & $\Lambda\gamma$  &  $100$  &  $70.40\pm 0.73\pm3.7$ &  $67.12\pm 0.69\pm 3.7$ &0.953 \\
$ \Sigma(1385)^+ $ & $\Lambda\pi^+$  &  $87\pm 1.5$  &$24.64\pm 0.39\pm 2.7$ &  $22.97\pm 0.32\pm 2.6$ &0.932 \\
$ \Xi^- $ & $\Lambda\pi^-$  &  $100$ &  $18.08\pm 0.18\pm 0.85$ &  $16.18\pm 0.16\pm 0.84$ &0.895 \\
$ \Xi(1530)^0 $ & $\Xi^-\pi^+$  &  $50$ &  $4.32\pm 0.070\pm 0.21$ &  $3.855\pm 0.062\pm 0.20$ &0.892 \\
$ \Omega^- $ & $\Lambda\,K^-$  &  $67.8\pm 0.7$ &  $0.995\pm 0.019\pm 0.048$ &  $0.887\pm 0.017\pm 0.047$ &0.891 \\
$ \Lambda_c^+ $ & $\pi^+K^-p$  &  $6.35\pm 0.33$ &  $157.76\pm 0.90\pm 8.0$ &  $141.79\pm 0.81\pm 7.8$ &0.899 \\
$ \Lambda_c(2595)^+ $ & $\Lambda_c^+\pi^+\pi^-$  &  $34.6 \pm 1.2$ &  $10.31\pm 0.011\pm 0.91$ & $10.157\pm 0.011\pm 0.92$ &0.985 \\
$ \Lambda_c(2625)^+ $ & $\Lambda_c^+\pi^+\pi^-$  &  $55.5 \pm 1.1$ &  $15.86\pm 0.12\pm 1.3$ & $15.37\pm 0.12\pm 1.3$ &0.969 \\
$ \Sigma_c(2455)^0 $ & $\Lambda_c^+\pi^-$  &  $100$ &  $8.419\pm 0.073\pm 1.2$ &  $7.963\pm 0.069\pm 1.1$ &0.946 \\
$ \Sigma_c(2520)^0 $ & $\Lambda_c^+\pi^-$  &  $100$ &  $8.31\pm 0.12\pm 1.3$ &  $7.77\pm 0.11\pm 1.3$ &0.935 \\
$\Omega_c^0$       &$\Omega^-\pi^+$ &   & $0.0153\pm 0.0020 \pm 0.00070$& $0.0130 \pm 0.0016 \pm 0.00060$& 0.850  \\
$\Xi_c^0$          &$ \Xi^-\pi^+  $ & & $0.376\pm 0.011\pm 0.013 $& $0.332\pm 0.010 \pm 0.013$& 0.880 \\
$\Xi_c^0$          &$ \Omega^-K^+ $ & & $ 0.110\pm 0.052 \pm 0.0038$& $0.097 \pm 0.046 \pm 0.0039$ & 0.880\\




\hline\hline
\end{tabular}
\end{table*}

\begin{figure*}
\includegraphics[scale=0.68]{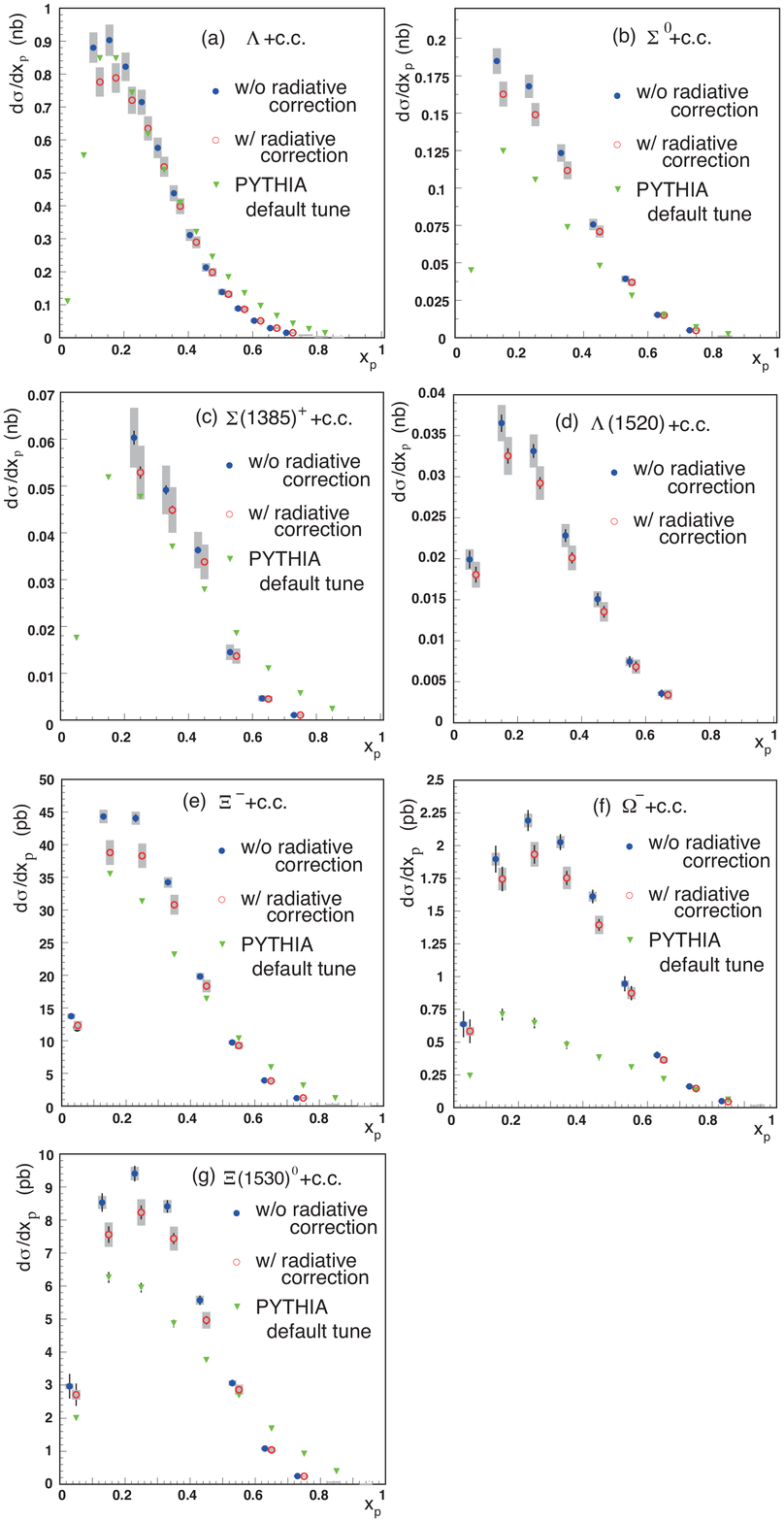}
\caption{Differential inclusive cross sections of hyperons with and without
radiative corrections. The closed circles are shifted slightly to the left for clarity.
The error bars and shaded boxes represent the statistical and
systematic uncertainties, respectively.
These distributions contain feed-down contributions from heavier particles.
Triangle points show predictions by PYTHIA with the default tune, where all radiative processes are
turned off, and the feed-down contributions are obtained using PYTHIA
predictions and branching
fractions given in Ref.~\cite{PDG2016}.}
\label{fig:cross_hyp1}
\end{figure*}

\begin{figure*}
\includegraphics[scale=0.9]{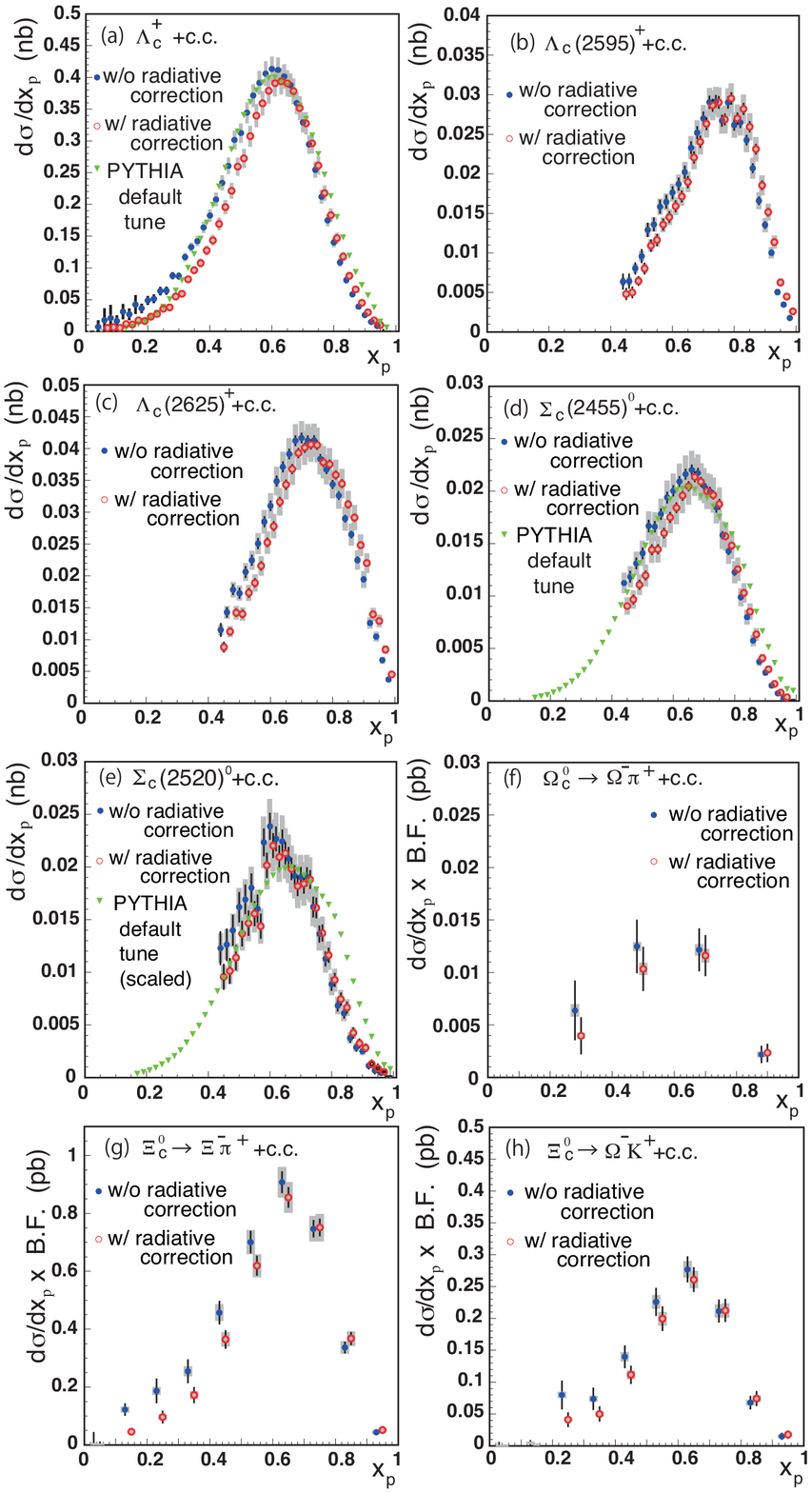}
\caption{Differential inclusive cross sections of charmed baryon production with and without
radiative corrections. The closed circles are shifted slightly to the left for clarity.
The error bars and shaded boxes represent the statistical and
systematic uncertainties, respectively.
These distributions contain feed-down contributions from heavier particles.
Triangle points show predictions by PYTHIA  with the default tune, 
where all radiative processes are
turned off, and the feed-down contributions are obtained using PYTHIA
 predictions and branching
fractions given in Ref.~\cite{PDG2016}.
Note that the prediction for $\Sigma_c(2520)^0$ is scaled by a factor of $0.5$.}
\label{fig:cross_charm}
\end{figure*}

The yields of hyperons and charmed baryons are obtained as a function of the 
scaled momentum, and corrections for 
reconstruction efficiencies are applied in each $x_p$ bin.
Reconstruction efficiencies are obtained using 
$e^+e^-\rightarrow q\bar{q}$ simulated events
that contain the particle of interest in the final state. 
Since we apply the reconstruction efficiency correction in each $x_p$
bin, the potential discrepancy of the momentum distributions 
between MC and real data is avoided.
The angular distributions of MC events are found to be consistent 
with those of real data.
The reconstruction efficiencies used in this analysis are shown 
in Appendix~\ref{Appendix_efficiency}.
The absolute branching fractions are obtained from Ref.~\cite{PDG2016} and
are used to calculate the production cross sections. The values used in this
analysis are listed in Table~\ref{tbl:cross_inclusive_radCor}.
The differential cross sections are shown in Figs.~\ref{fig:cross_hyp1}
and \ref{fig:cross_charm}.
We note that these cross sections contain feed-down contributions from
higher resonances (inclusive cross sections). 

The correction factor due to the initial state radiation (ISR) and the
vacuum polarization of the virtual gauge
bosons in $e^+e^-$ annihilation is studied by PYTHIA~\cite{PYTHIA} by comparison
between the cross sections computed with and without inclusion of the ISR
and the vacuum polarization. Both virtual gamma and $Z^0$
exchanges including the interference between them are taken into
account as the PYTHIA default. The effect of
the final state radiation (FSR) from charged particles is investigated
using PHOTOS program~\cite{PHOTOS} and we confirm that the FSR gives only
negligible effect to present cross sections.
For each particle species, MC events are generated with and without
the ISR and the vacuum polarization effects and consequently $d\sigma/dx_p$ distributions are
obtained. Using PYTHIA, the total hadronic cross sections with and
without inclusion of the ISR and the vacuum polarization are calculated to be 3.3~nb and 2.96~nb
respectively. We get the correction factors in each $x_p$ bin by taking
the ratio between $d\sigma/dx_p$ for with and without radiative
correction terms by scaling the ratio according to the calculated total
hadronic cross sections.
{In the case of an ISR event the CM energy of the $e^+e^-$ annihilation
process reduces and the true and reconstructed $x_p$ will be different. The
ratio of $x_p$ distributions without ISR over ISR is taken to correct the
differential cross sections.}
The differential cross sections before and after the correction are
shown in Figs.~\ref{fig:cross_hyp1} and \ref{fig:cross_charm}.

Figures~\ref{fig:cross_hyp1}(a)-(d) show the differential cross sections
for $S=-1$ hyperons.
In the low $x_p$ and high $x_p$ regions, 
the signals of hyperons are not significant due to the small
production cross sections and large number of background events.
We obtain total cross sections over the entire $x_p$ region by 
utilizing a third-order Hermite interpolation 
describing the behavior in the measured $x_p$ range, 
where we assumed that the cross section is zero at $x_p=0$ and $x_p=1$.
We obtain total cross sections over the entire $x_p$ region by 
utilizing a third-order Hermite interpolation 
describing the behavior in the measured $x_p$ range, 
where we assumed that the cross section is zero at $x_p=0$ and $x_p=1$.
The estimated contributions from the unmeasured $x_p$ regions are 19\%,
15\%, and 49\% of the contributions from the measured regions for $\Lambda$, $\Sigma^0$, 
$\Sigma(1385)^+$, and $\Lambda(1520)$, respectively.
We also estimate the contributions from the unmeasured regions by
assuming the PYTHIA spectrum shapes. The differences between the two estimations
are typically 20-30\% and assigned to the systematic errors for the extrapolation of  
the cross sections.
For $S=-2$ and $-3$ hyperons, the cross sections are measured in the entire
$x_p$ region.


The differential cross sections for charmed baryons after the correction for the 
reconstruction efficiency and the branching fractions 
 are shown in Fig.~\ref{fig:cross_charm}.
Here, we utilize the world-average absolute branching fraction of
${\cal B}(\Lambda_c^+\rightarrow K^-\pi^+ p)=(6.35 \pm 0.33)$\%~\cite{PDG2016}.
The branching fractions of $\Lambda_c(2595)^+\rightarrow\Lambda_c^+\pi^+\pi^-$ 
and $\Lambda_c(2625)^+\rightarrow \Lambda_c^+\pi^+\pi^-$
are determined to be $0.346\pm 0.012\ ({\rm syst.})$ and 
$0.555 \pm 0.011\ {\rm (syst.)}$, utilizing the model by 
Cho~\cite{ChoPRD50} and accounting for the mass difference of the
charged and neutral pion. 
Details are described in Appendix~\ref{Appendix_feed-down}.
Since the absolute branching fractions of $\Omega_c^0\rightarrow
\Omega^-\pi^+$, $\Xi_c^0\rightarrow \Xi^-\pi^+$ and 
$\Xi_c^0\rightarrow \Omega^-K^+$ are unknown, the cross section times 
the branching fraction are plotted in Figs.~\ref{fig:cross_charm}(f)-(h).
%
The cross sections for $\Lambda_c(2595)^+$, $\Lambda_c(2625)^+$,
$\Sigma_c(2455)^0$, and $\Sigma_c(2520)^0$
in the $0.44<x_p<1$ region after the radiative correction
are $(9.60 \pm 0.08)$~pb, $(11.39 \pm 0.07)$~pb, 
$(6.34 \pm 0.04) $~pb, and $(6.07 \pm 0.08) $~pb, respectively.
Clearly, the production cross sections for $\Lambda_c^+$ excited
states are significantly higher than those for $\Sigma_c^0$ baryons in the
measured $x_p$ region without the extrapolation to the whole $x_p$ region.
We note that the radiative correction factors are consistent within 4\%
for these particles and are not the source of the difference of the production
cross sections.
We obtain cross sections of excited $\Lambda_c^+$ and $\Sigma_c^0$ states
in the entire $x_p$ region utilizing the $x_p$ dependence of 
cross sections obtained 
from MC using the Lund model~\cite{Andersson_PhysRept}. 
The correction factors {for extrapolating from the measured $x_p$ region to 
the entire $x_p$ region} are small: 1.07, 1.07,
1.16, and 1.18 for $\Lambda_c(2595)$, $\Lambda_c(2625)$, 
$\Sigma_c(2455)^0$, 
and $\Sigma_c(2520)^0$, respectively.
We obtain alternate correction factors using fragmentation models---BCFY~\cite{BCFY}, Bowler~\cite{Bowler}, Peterson~\cite{Peterson}, 
and KLP-B~\cite{KLPB}---and 
take the deviations of about 5 to 12\% as the systematic uncertainty.

%

Table~\ref{tbl:cross_inclusive_radCor} shows cross sections before and after
the radiative corrections.
The correction factors are consistent for hyperons; however, larger
correction factors by about 5\% are obtained for the excited $\Lambda_c^+$ baryons
than for $\Sigma_c^0$ baryons.
The $d\sigma/dx_p$ distribution is harder for the excited
$\Lambda_c^+$ baryons, as shown in Fig.~\ref{fig:cross_charm}, and
the cross sections in the high-$x_p$ (low-$x_p$) region are increased
(reduced) due to the radiative cross sections.
As a result, we have larger correction factors for the excited $\Lambda_c^+$ baryons.
The systematic uncertainties are discussed in Sec.~\ref{sect:syst}.

Triangle points in Figs.~\ref{fig:cross_hyp1} and \ref{fig:cross_charm}
show predictions by PYTHIA with default parameters, where all radiative processes are
turned off. The feed-down contributions are obtained using PYTHIA
predictions and branching fractions given 
in Ref.~\cite{PDG2016}.
Note that the prediction for $\Sigma_c(2520)^0$ overestimates the experimental
data, and we scaled the predicted values by a factor of $0.5$.

\subsection{Systematic uncertainties}
\label{sect:syst}

\begin{table*}[ht]
 \caption{Systematic uncertainties (\%) for the total cross section of hyperons and charmed strange baryons.
The $\Lambda$ detection efficiency includes proton and pion identification efficiencies. 
The symbols of ``-'' and $\bigcirc$ mean that the uncertainty is much smaller than the statistical fluctuation and
that the uncertainty is not taken into account, respectively.
}
  \label{tbl:syst_s-1}
\begin{tabular}{l c c c c c c c c c c} \hline\hline
 Source               & $\Lambda$   & $\Sigma^0$  & $\Sigma(1385)^+$ & $\Lambda(1520)$ & $\Xi^-$ & $\Omega^-$ & $\Xi(1530)^0$ & $\Xi_c^0$ in & $\Xi_c^0$ in  &$\Omega_c^0$\\
                      &             &             &                  &                 &           &            &              &  $\Xi^-\pi^+$ & $\Omega^-K^+$ & \\ \hline
Track reconstruction  &  0.70       & 0.70        & 1.1              & 0.70            &  1.1      & 1.1        & 1.4          &  1.4          & 1.4           & 1.4 \\ 
$\Lambda$ detection   &  2.8        & 2.8         & 2.8              & $\bigcirc$      & 3.3       & 3.2        & 3.0          & 3.3           & 3.2           & 3.2\\ 
$\gamma$ detection    &  $\bigcirc$ & 2.0         & $\bigcirc$       & $\bigcirc$      & $\bigcirc$&$\bigcirc$  & $\bigcirc$   & $\bigcirc$  & $\bigcirc$    & $\bigcirc$\\
Particle ID           &  $\bigcirc$ & $\bigcirc$  & 1.3              & 1.1             & 1.1     & 1.1        & 1.5          & 1.4           & 1.1           & 1.1 \\
MC statistics         &  0.10       & 0.75        & 2.0              & 1.2             & 0.10    & 0.95       & 0.39         & 0.22          & 0.39          & 0.55 \\
Signal shape          & $\bigcirc$  & 1.4         & 0.57             & 2.8             & 0.2     & 0.6        & 2.0          & 3.4           & 0.2           & 1.2  \\
Background estimation & $\bigcirc$  &  -          & 2.2              & 5.4             & 1.0      & 1.0        & 1.0          & 1.0           & 1.0           &1.0\\
Experimental period   &   -         &  -          & -                &  -              &   -     &  -         & -            &   -           &  -            & - \\
Baryon anti-baryon    &   -         &  -          & -                &  -              &   -     &  -         & -            &   -           &  -            & -  \\
Impact parameter      &   -         &  -          & -                &  -              &   -     &  -         & -            &   -           &  -            & - \\
Extrapolation of $d\sigma/dx_p$& 3.8& 2.1         & 9.4              & 0.96            &    -    &  -         &  -           &   -           &  -            & - \\
Radiative correction   &2.3 &2.3 &2.3 &2.3 &2.3      & 2.3        &2.3           &2.3 &2.3 &2.3 \\ 
Luminosity measurement &1.4 &1.4 &1.4 &1.4 &1.4      & 1.4        &1.4            &1.4 &1.4 &1.4\\
Total                  & {\bf 5.5} & {\bf 5.1}    & {\bf 11}         & {\bf 6.9}       & {\bf 4.6} & {\bf 4.7} & {\bf 5.0} & {\bf 5.8} & {\bf 4.7}  & {\bf 4.7} \\
\hline\hline
\end{tabular}
\end{table*}





\begin{table*}[ht]
 \caption{Systematic uncertainties (\%)  for the total cross section
 of charmed baryons.
The symbols of ``-'' and $\bigcirc$ mean that the uncertainty is much smaller than the statistical fluctuation and
that the uncertainty is not taken into account, respectively.
}
  \label{tbl:syst_charm}
\begin{tabular}{l c c c c c} \hline\hline
Source                          & $\Lambda_c^+$     &$\Lambda_c(2595)^+$ &$\Lambda_c(2625)^+$ &$\Sigma_c^0$ & $\Sigma_c(2520)^0$\\\hline
Track reconstruction            & 1.1               & 1.8                & 1.8                 & 1.4        & 1.4    \\
Particle ID                     & 2.0               & 3.9                & 4.0                 & 5.0        & 1.4    \\
MC statistics                   &0.27               & 0.10               & 0.30                & 0.10       & 0.14   \\
Signal shape                    &$\bigcirc$         & 2.8                & 1.3                 & 2.2        & 1.5    \\
Background  estimation          &$\bigcirc$         & 2.0                & 2.3                 & 1.0        & 7.5    \\
Experimental period             &1.8                & -                  & -                   &2.5         &5.9     \\
Baryon anti-baryon              &1.5                & -                  & -                   & -          &  -     \\
Impact parameter                & 2.2               & -                  & -                   & -          & -      \\
$B$-meson decay                 & -                 & 3.7                & 2.6                 &3.3         &0.6     \\
Extrapolation of $d\sigma/dx_p$ & -                 & 5.7                & 5.6                 &11          &12      \\
Radiative correction   &2.3 &2.3 &2.3 &2.3 &2.3 \\ 
Luminosity  measurement &1.4 &1.4 &1.4 &1.4 &1.4 \\
Total       & {\bf 4.8}      & {\bf 9.1}            & {\bf 8.4} & {\bf 13} & {\bf 16} \\
\hline\hline
\end{tabular}
\end{table*}

The sources of systematic uncertainties are summarized in
Tables~\ref{tbl:syst_s-1} and \ref{tbl:syst_charm}.
The uncertainties due to the reconstruction efficiency of charged particles
and the $\Lambda$ selection including particle
identification (particle ID) are estimated by comparing the efficiencies
in real data and MC.
The systematic uncertainty of photon detection efficiency for
$\Sigma^0\rightarrow \Lambda \gamma$ decay is estimated
to be 2\% from a radiative Bhabha sample.
The uncertainties of the particle ID for kaons, pions, and protons are estimated
by comparing the efficiencies in real data and MC, where 
$D^0\rightarrow K^-\pi^+$ events and $\Lambda\rightarrow p\pi^-$ events are
used for kaon (pion) selection and proton selection, respectively.
The uncertainties of the reconstruction efficiency due to the statistical fluctuation
of the MC data are taken as systematic uncertainties.

The signal shapes for $\Sigma^0$, $\Xi^-$, $\Omega^-$, 
$\Xi_c$, and $\Omega_c$ are
assumed as double Gaussian. First, we confirm that the background shape
is stable by changing the signal shape.
We compare the signal yield with the one obtained by 
subtracting the background contribution from the total
number of events, and take the difference as the systematic uncertainty
due to the background shape.
For excited particles, 
we estimate the systematic uncertainty due to the signal shape by fixing
the resolution parameter of Voigtian function to the value obtained by MC.
The yields of the ground-state $\Lambda$ and $\Lambda_c$ are obtained by
sideband subtraction, and the systematic uncertainties due to the signal
shape are not taken into account.

The uncertainty due to the background estimation for hyperons and
charmed strange baryons is determined by 
utilizing a higher order polynomial to describe the background contribution 
and then redetermining the signal yield.
For the yield estimation of excited charmed baryons, 
the background shape described by the threshold function is compared with
the background shape obtained by MC, in which the threshold function is
given by $a(m-m_{0})^b\exp(c(m-m_{0}))$, where $m$ is the invariant
mass, $m_0$ is the threshold value, and $a$, $b$ and $c$ are fit parameters.
The differences of the obtained signal yields are taken 
as the systematic uncertainty.
The yields of the $\Lambda$ and $\Lambda_c^+$ baryons are
obtained by sideband subtraction. Because the uncertainty of the background estimation 
is included in the statistical uncertainties here, 
this uncertainty is not taken as a systematic uncertainty.

To evaluate other sources of systematic uncertainties, the cross sections
are compared using subsets of the data: events recorded in the different experimental
periods, or the baryon \textit{vs.}  anti-baryon samples.
In addition, the cross sections are compared by changing the event-selection
criteria: impact-parameter requirements for tracks, 
or the $x_p$ threshold to eliminate the $B$-meson decay
contribution for excited $\Lambda_c$ and $\Sigma_c$ baryons.
If these differences are larger than the statistical fluctuation,
we take them as systematic uncertainties.

We estimate the uncertainties due to the extrapolation to
the whole $x_p$ range for $S=-1$ hyperons using, the $d\sigma/dx_p$ 
distribution of MC events for the extrapolation, 
which are generated using Lund fragmentation model.
We compare the results of the extrapolation using all measured points 
and only the lowest $x_p$ data (where the feed-down contribution is
large); and the largest discrepancy is taken as the systematic
uncertainty.

The systematic uncertainty due to the radiative correction is estimated
using PYTHIA. However, because we apply radiative corrections in each 
$x_p$ bin, we expect the dependence of the correction factors on the 
fragmentation model to be reduced.
The largest difference of the correction factors for different PYTHIA tunes,
which were described in Ref.~\cite{Belle_pair_cross_section}, is 
2.1\%, and is taken as a 
systematic uncertainty that is common for all $x_p$ bins.
An additional uncertainty due to the accuracy of radiative effects in
the generator is estimated to be 1\%~\cite{Kleiss_Z_LEP}, 
and is taken as a systematic uncertainty.

The uncertainty due to the luminosity measurement (1.4\%) is common for
all particles.
The $x_p$ dependence of the systematic uncertainty is found to be
less than 0.4\% and is negligible for all particles.

\subsection{Direct cross sections}

\begin{table*}[htb]
  \caption{Direct cross sections after the feed-down subtraction, and
the fraction of the direct cross sections with respect to the radiative-corrected
 cross sections.
Direct cross sections predicted by PYTHIA with default 
parameters are listed for the positive-parity baryons, where radiative processes are turned off.
The masses and spins used in
Figs.~\ref{fig:cross_mass_fit_hyp} and \ref{fig:cross_mass_fit_charm} 
are itemized.
}
  \label{tbl:cross_inclusive_direct}
\begin{tabular}{ l c c c  c c c c c c } \hline\hline
Particle & Mass         & Spin &  Direct       & Fraction  & PYTHIA\\
         &  (MeV/$c^2$) &      &  cross        &           & prediction\\
         &              &      &  section (pb) &           & (pb)\\ \hline
$\Lambda $ & 1115.6 & 1/2 & $91.2\pm 2.1\pm 22 $ &0.32 &$87.7\pm0.4$\\
$\Lambda(1520) $ & 1519.5 & 3/2 & $9.68\pm 0.75\pm 0.26 $ &0.73 &\\
$\Sigma^0 $ & 1192.6 & 1/2 & $52.28\pm 0.66\pm 3.8 $ &0.83 & $36.1\pm0.3$\\
$\Sigma(1385)^+ $ & 1382.8 & 3/2 & $18.39\pm 0.35\pm 2.8 $ &0.83 & $19.8\pm0.2$\\
$\Xi^- $ & 1321.4 & 1/2 & $11.25\pm 0.17\pm 0.33 $ &0.7 & $10.8\pm0.1$\\
$\Xi(1530)^0 $ & 1531.8 & 3/2 & $3.855\pm 0.062\pm 0.22 $ &1.0 &$2.58\pm0.07$\\
$\Omega^- $ & 1672.4 & 1/2 & $0.887\pm 0.017\pm 0.047 $ &1.0 & $0.32\pm0.02$\\
$\Lambda_c^+ $ & 2286.4 & 1/2 & $67.6\pm 1.5\pm 9.1 $ &0.48 &$85.8\pm0.3  $\\
$\Lambda_c(2595)^+ $ & 2592.2 & 1/2 & $10.157\pm 0.011\pm 0.92 $ &1.0 & \\
$\Lambda_c(2625)^+ $ & 2628.1 & 3/2 & $15.367\pm 0.116\pm 1.3 $ &1.0 & \\
$\Sigma_c(2455)^0 $ & 2453.7 & 1/2 & $6.697\pm 0.069\pm 1.2 $ &0.84 & $8.5\pm0.1$\\
$\Sigma_c(2520)^0 $ & 2518.8 & 3/2 & $7.77\pm 0.11\pm 1.3 $ &1.0 &$16.6\pm0.1 $\\
\hline\hline
\end{tabular}
\end{table*}

Our motivation is to search for the enhancement or the reduction of the production 
cross sections of certain baryons and to discuss their internal structures, as
 described in Sec.~\ref{sec:intro}. 
For this purpose, the subtraction of feed-down from heavier particles is quite important
since the amount of this feed-down is determined by the production cross sections
of mother particles and the branching fractions, which are not related to
the internal structure of the baryon of interest.
Table~\ref{tbl:cross_inclusive_direct} shows the inclusive cross sections
after the feed-down subtraction (direct cross
section) and their fraction of the cross sections after the radiative correction.
The branching fractions and feed-down contributions are
summarized in Appendix~\ref{Appendix_feed-down}. 
We use the world-average branching fractions in Ref.~\cite{PDG2016}.
We should note that the cited list may be incomplete,
\textit{i.e.,}  we may have additional feed-down contributions.
Such contributions are expected to be small, and should be 
subtracted when the branching fractions are measured in the future.
While the calculation of the feed-down contributions, the 
same production rates are assumed for iso-spin partners 
($\Sigma(1385)$, $\Xi$, $\Xi(1530)$, $\Sigma_c(2455)$, $\Sigma_c(2520)$).
The branching fraction of ${\cal B}(\Lambda_c(2595)^+ \rightarrow \Lambda_c^+\pi^+\pi^-)$
is obtained to be $0.346 \pm 0.012 \ ({\rm syst.})$ using Cho's function~\cite{ChoPRD50}
with the parameter obtained by CDF~\cite{CDF_PRD84012003}.
More details are described in Appendix~\ref{Appendix_feed-down}.

The systematic uncertainties for the feed-down contribution are
calculated using those for the inclusive cross sections of mother
particles, and we use the quadratic sum for the systematic 
uncertainty of the direct cross section.
The uncertainty of the luminosity measurement is common to all baryons,
and, in order to avoid double counting, 
we add the uncertainty due to the luminosity to the cross sections after the
feed-down subtraction.
The uncertainties for the branching fractions are taken as the
systematic uncertainties for the direct cross sections.

\section{Results and discussion}
\label{sec:Results}

\subsection{Scaled momentum distributions}
\label{sect:discuss_momentum}

We discuss the differential cross sections first.
The open circles in Figs.~\ref{fig:cross_hyp1} and \ref{fig:cross_charm}
show $d\sigma/dx_p$ for hyperons and charmed baryons after the
radiative correction.
{The differential production cross sections of hyperons peak in the 
small $x_p$ region compared to those of charmed baryons. This behavior
suggests that, at energies near $\sqrt{s}=10.5$~GeV, $s\bar{s}$ pairs 
that lead to hyperons are created mainly in the soft processes in the 
later stage of the fragmentation rather than in the hard processes of 
prompt $s\bar{s}$ creation from the initial virtual photon. 
The $d\sigma/dx_p$ distribution of charmed baryons show peaks in the high
$x_p$ region, since  $c\bar{c}$ pairs are created predominantly in the prompt
$e^+e^-$ collision, and charmed baryons carry a large fraction of the initial beam energy.}

The peak cross section of the hyperons occurs below $x_p=0.2$ and is 
consistent for all $S=-1$ hyperons. 
The $d\sigma/dx_p$ distributions for $S=-2, -3$ hyperons 
(Figs.~\ref{fig:cross_hyp1}(e)-(g)) exhibit peaks at
slightly higher $x_p$ ($x_p > 0.2$) than for $S=-1$ hyperons.
Since the strange quark is heavier than the up or down quark, the energy
necessary to create an $S=-2$ hyperon is larger than an $S=-1$ hyperon, and
$S=-2$ hyperons may be produced in a rather harder process than $S=-1$ ones.

The distribution for the $\Lambda_c(2286)^+$ peaks at
$x_p=0.64$, and that for the $\Sigma_c(2455)^0$ peaks at $x_p=0.68$. 
The peak position for the $\Sigma_c(2520)^0$ is not
determined clearly due to the statistical fluctuations.
The distributions for the $\Lambda_c(2595)^+$ and
the $\Lambda_c(2625)^+$ show peak structures at significantly higher $x_p$
($x_p=0.78$). 
The peak position for the $\Xi_c(2470)^0$ is around $x_p=0.65$, which is
consistent with the $\Lambda_c(2286)^+$ and the $\Sigma_c(2455)^0$.

\begin{figure*}
\includegraphics[scale=0.7]{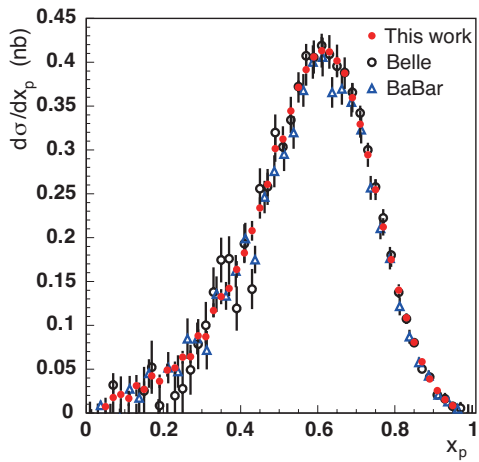}
\caption{
The differential cross sections of $\Lambda_c^+$ production before the radiative 
correction, where the absolute branching fraction of
${\cal B}(\Lambda_c^+\rightarrow \pi^+K^-p)=0.0635$ is used to normalize
the previous results~\cite{BaBar_Lc,Belle_Charm_Seuster} for the comparison.
To scale the multiplicity measurement by BaBar, the total hadronic cross section of 
3.3~nb is utilized.
The error bars represent the sum in quadrature of the statistical and systematic uncertainties; note that
Belle's previous measurement contains an additional uncertainty of 35\% for the normalization.}
\label{fig:invm_Lc2286_comp}
\end{figure*}

\subsection{Comparison of inclusive cross sections with previous results}

Table~\ref{tbl:comparison_inclusive} shows a comparison with previous measurements,
where for hyperons, we use the hadron multiplicities that were measured by
ARGUS~\cite{ARGUS_hyperon,ARGUS_L1520}, since
the statistics of other results are quite limited.
For charmed baryons, we utilize the measurement of $\Lambda_c^+$ production by BaBar~\cite{BaBar_Lc} 
and the ratios of production rates of excited particles relative to the $\Lambda_c^+$ 
measured by CLEO~\cite{CLEO_Lc2595_2625,CLEO_Sc2455,CLEO_Sc2520} and
ARGUS~\cite{ARGUS_Lc2595,ARGUS_Lc2625}.
For the comparison, we utilize the world-average 
absolute branching fraction of ${\cal B}(\Lambda_c^+\rightarrow
\pi^+K^-p)=0.0635$~\cite{PDG2016} to normalize
the previous results of charmed baryons. 
Since previous measurements report cross sections without the radiative
correction, we compare our results for the visible cross section. 
The total hadronic cross section of 3.3~nb~\cite{CLEO_Hadronic} is used to 
normalize hadron multiplicities to cross sections.

The differential cross section of $\Lambda_c^+$ production before the radiative
correction is compared with the prior measurements by BaBar~\cite{BaBar_Lc} 
and Belle\cite{Belle_Charm_Seuster} as shown in Fig.~\ref{fig:invm_Lc2286_comp}.
For comparison, the absolute branching fraction of
${\cal B}(\Lambda_c^+\rightarrow \pi^+K^-p)=0.0635$~\cite{PDG2016} is used
to rescale both of the BaBar and Belle measurements for this figure.
To scale the multiplicity measurement by BaBar, the total hadronic cross section of 
3.3~nb is utilized.
Our result is consistent with these previous measurements.

We observe that the production cross sections of hyperons are consistent with
previous measurements but with much higher precision.
Here, it is noted that the statistics of the 
$\Lambda(1520)$ in the ARGUS result is quite limited.
Their result is slightly larger than this work; however, is consistent
within 2.0$\sigma$ due to the large uncertainty on their measurement.
The production rate of the $\Lambda_c(2595)^+$ by this work is larger than the CLEO result; 
the corresponding ARGUS result~\cite{ARGUS_Lc2595} is consistent with ours, 
but contains a large uncertainty due
to the extrapolation to the whole $x_p$ region.
ARGUS reported a more precise  production cross section for $x_p>0.7$ 
of $(8.0\pm 2.3 \,({\rm stat.}) \pm 1.7 \,({\rm syst.}))$~pb, 
which is consistent with our result of $(7.34 \pm 0.06 \ ({\rm stat.}))$~pb.
The production rate of the $\Lambda_c(2625)^+$ in this work is significantly larger 
than the CLEO result. The ratio of production rates of the $\Lambda_c(2625)^+$
to the $\Lambda_c(2595)^+$ is about 1.3 and is consistent with this work.
The result obtained by ARGUS is slightly larger than
the CLEO result and closer to our result.

\begin{table*}[htb]
  \caption{Comparison of visible cross sections with previous measurements.
The first and second errors represent the statistical and systematic uncertainties,
respectively.}
  \label{tbl:comparison_inclusive}
\begin{tabular}{ l c c c} \hline\hline
Particle & Visible cross section & Visible cross section by & References for\\
         & by this work (pb)     & previous measurements (pb) & previous measurements \\ \hline
$\Lambda$           & $308.8\pm 0.37\pm 17$     & $306\pm 10\pm 26$ & \cite{ARGUS_hyperon}\\
$\Lambda(1520)$     & $14.32\pm0.23\pm 1.0$     & $26.6\pm 5.7\pm 4.3$ & \cite{ARGUS_L1520} \\
$\Sigma^0$          & $70.40\pm 0.73\pm3.7$     & $76\pm 22 \pm16$ & \cite{ARGUS_hyperon} \\
$\Sigma(1385)^+$    & $24.64\pm 0.39\pm 2.7$    & $17.1\pm 3.1\pm 3.1$ & \cite{ARGUS_hyperon} \\
$\Xi^-$             & $18.08\pm 0.18\pm 0.85$   & $22\pm 2\pm 2$ & \cite{ARGUS_hyperon} \\
$\Xi(1530)^0$       & $4.32\pm 0.070\pm 0.21$    & $4.9\pm 1.7\pm 0.77$ & \cite{ARGUS_hyperon} \\
$\Omega^-$          & $0.995\pm 0.019\pm 0.048$ & $2.4\pm 1.2\pm 0.43$ & \cite{ARGUS_hyperon} \\
$\Lambda_c^+$       & $157.76\pm 0.90\pm 8.0$      & $148.9\pm 1.8\pm 1.6$ & \cite{BaBar_Lc} \\
$\Lambda_c(2595)^+$ & $10.31\pm 0.011\pm 0.91$  & $6.1\pm 1.0\pm 1.3$ & \cite{CLEO_Lc2595_2625} \\
$\Lambda_c(2595)^+$ &                           & $11.2^{+10.8}_{-5.8}\pm 8.3$ & \cite{ARGUS_Lc2595} \\
$\Lambda_c(2625)^+$ & $15.86\pm 0.12\pm 1.3$    & $7.80\pm 0.76\pm 0.62$ & \cite{CLEO_Lc2595_2625} \\
$\Lambda_c(2625)^+$ &                           & $10.8\pm 2.4\pm 2.8$ & \cite{ARGUS_Lc2625} \\
$\Sigma_c^0$        & $8.419\pm 0.073\pm 1.2$   & $8.9\pm 1.5\pm 2.5$ & \cite{CLEO_Sc2455} \\
$\Sigma_c(2520)^0$  & $8.31\pm 0.12\pm 1.3$     & $9.5\pm 1.1\pm 2.4$ & \cite{CLEO_Sc2520} \\
\hline\hline
\end{tabular}
\end{table*}

%
 

\subsection{Mass dependence of direct production cross sections}

\begin{figure}[ht]
 \includegraphics[scale=0.4]{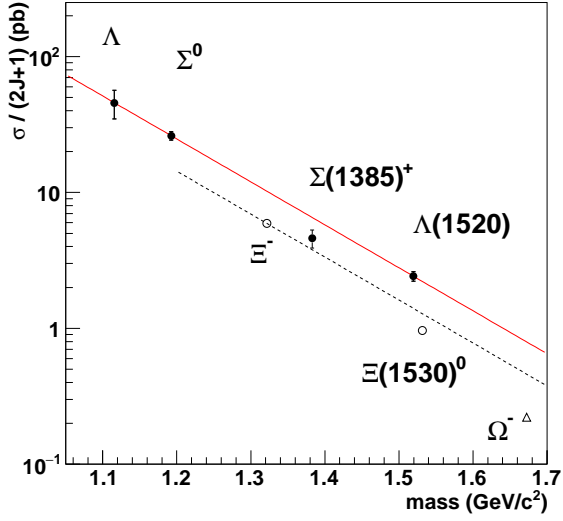}
   \caption{Scaled direct production cross section as a function of mass of hyperons.
 $S=-1,-2,-3$ hyperons are shown with filled circles, open circles and
 a triangle, respectively.
The solid line shows the fit result using an exponential function
(Eq.\ref{eq:cross_mass}) 
 for $S=-1$ hyperons except for $\Sigma(1385)^+$. The dashed line
shows an exponential curve with the same slope parameter as $S=-1$ hyperons,
which is normalized to the production cross section of $\Xi^-$.
}
   \label{fig:cross_mass_fit_hyp}
\end{figure}

\begin{figure}[ht]
 \includegraphics[scale=0.4]{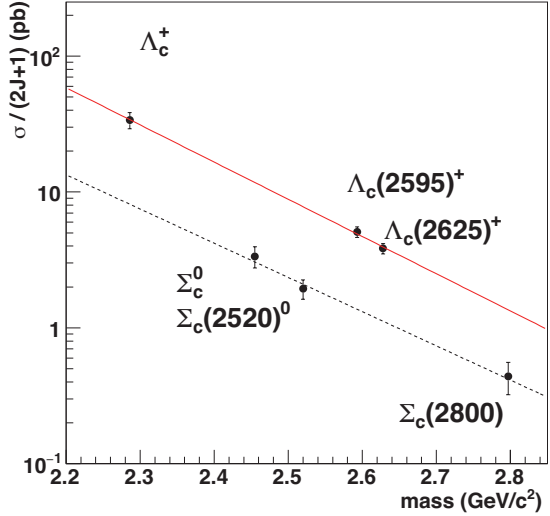}
   \caption{Scaled direct production cross section as a function of mass of charmed 
baryons.
The solid and dashed lines show the fit results using exponential functions
(Eq.\ref{eq:cross_mass}) 
 for $\Lambda_c$ baryons and $\Sigma_c$ baryons, respectively.
}
   \label{fig:cross_mass_fit_charm}
\end{figure}

We divide the direct production cross sections by the number of spin states
$(2J+1)$ and plot these as a function of baryon masses
(Figs.~\ref{fig:cross_mass_fit_hyp} and \ref{fig:cross_mass_fit_charm}).
The error bars represent the sum in quadrature of the statistical and systematic
uncertainties.
In Fig.~\ref{fig:cross_mass_fit_hyp}, the production cross sections of $S=-1$ hyperons
show an exponential dependence on the mass except for the
$\Sigma(1385)^+$.
We fit the production cross sections of $S=-1$ hyperons except for the $\Sigma(1385)^+$
using an exponential function, 
\begin{equation}f(m)=a_0\exp(a_1 m), \label{eq:cross_mass} \end{equation} where $m$ is the mass
of the particle and $a_0$ and $ a_1$ are fit parameters; we obtain 
$a_0=(1.6\pm 0.7)\times 10^5$~pb, $a_1=(-7.3\pm 0.3)/$(GeV/c$^2$). 
Due to the large uncertainty on the $\Lambda$ hyperon,
the $\chi^2/ndf$ value is very small.

We do not observe the enhancements of the direct cross sections of 
$\Lambda$ and $\Lambda(1520)$ that
were discussed in Refs.~\cite{Jaffe_Exotica,Wilczek_diquark} because
they used data of inclusive production, which contain
large feed-down contributions from heavier particles.
The scaled direct cross sections for $\Lambda$, $\Sigma^0$ and $\Lambda(1520)$
follow an exponential mass dependence with a common slope parameter. 
{The scaled direct cross section for $\Sigma(1385)^+$ is 
smaller than the predicted value of the exponential curve at $m=1.382$~GeV/c$^2$
by 30~\% with the statistical significance of 2.8$\sigma$},
as was reported by ARGUS~\cite{ARGUS_hyperon}. 
We found that the fit including the $\Sigma(1385)^+$ results in the
deviation of 2.2$\sigma$. 
{As already mentioned, the predicted production rate of the diquark model
is smaller than that of the popcorn model by 30~\%. However, 
these predictions include feed-down contributions,
and predictions for the direct production cross sections are desired.}

Since the mass of a strange quark is heavier than of an up or down quark, 
the probability of the $s\bar{s}$ pair creation is expected to be smaller
than that of the non-strange quark pair creation.
Indeed, $S=-2$ and $-3$ hyperons have significantly smaller
production cross sections compared to $S=-1$ hyperons, which are likely due to
the suppression of $s\bar{s}$ pair creation in the fragmentation process.
Despite the mass difference between strange and lighter quarks, 
one may expect the same mechanism to form a baryon between $S=-1$ and $S=-2$ hyperons.
The dashed line in Fig.~\ref{fig:cross_mass_fit_hyp} shows
an exponential curve with the same slope parameter as $S=-1$ hyperons,
which is normalized to the production cross section of $\Xi^-$.
Clearly,  the production cross section of the $\Xi(1530)^0$ is suppressed with
respect to this curve. This may be due to the decuplet suppression
noted in the $\Sigma(1385)^+$ case.
The production cross section for the $S=-3$ hyperon, $\Omega^-$, shows further
suppression for the creation of an additional strange quark.

The results for charmed baryons are shown in 
Fig.~\ref{fig:cross_mass_fit_charm}.
The production cross section of the $\Sigma_c(2800)$ measured by Belle~\cite{Belle_Sigmac2800}
is shown in the same figure, where
we utilize the weighted average of cross sections for the three charged states,
and assume that the $\Lambda_c^+\pi$ decay mode dominates over the others. 
In Ref.~\cite{Belle_Sigmac2800}, the spin-parity is 
tentatively assigned as $J^P=3/2^-$, so we use a spin of $3/2$ for
this state.

The prompt production of a $q\bar{q}$ pair from $e^+e^-$ annihilation couples to
the charge of quarks. If the center-of-mass energy of $e^+e^-$ is high 
compared to the mass of the charm quarks, the production rates of charm quarks
become consistent with those of up quarks. 
Indeed, near the $\Upsilon(4S)$ energy, 
the production cross section of the $\Lambda_c^+$ ground state is much higher than
the exponential curve of hyperons {in Fig.}~\ref{fig:cross_mass_fit_hyp} 
extended to the mass of charmed baryons.
The production mechanism of charmed baryons differs from that of
hyperons.
For charmed baryons, a $c\bar{c}$ pair is created from the prompt 
$e^+e^-$ annihilation and picks
up two light quarks to form a charmed baryon. {Since this process occurs
in the early stage of the fragmentation process where the number of quarks
are few, the probability to form a charmed baryon from uncorrelated quarks
is smaller than that from diquark and anti-diquark production.
In addition to the production mechanism, we note that the diquark correlation
in the charmed baryons are stronger than that in hyperons due to the
heavy charm quark mass as discussed in Sec.}~\ref{sec:intro}.
{Although these interpretations are model dependent}, we can expect that 
the production cross sections of charmed baryons 
are related to the production cross sections of diquarks.

The production cross sections of $\Sigma_c$ baryons
are smaller than those of excited $\Lambda_c^+$ by a factor of about three,
in contrast to hyperons where $\Lambda$ and $\Sigma$ resonances lie
on a common exponential curve.
This suppression is already seen in the cross section in the $0.4<x_p<1$
region, and is not due simply to the extrapolation by the fragmentation models.

Table~\ref{tbl:cross_inclusive_direct} shows the direct cross sections
predicted by PYTHIA6.2 using default parameters.
Note that PYTHIA can not produce negative-parity baryons.
The predicted cross sections are consistent with the experimental measurements
for hyperons except for $\Sigma^0$, $\Xi(1530)^0$ and $\Omega^-$.
However, for charmed baryons, PYTHIA overestimate the experimental results.
Since theoretical predictions for the production rates of charmed
baryons are not available, we analyze our data assuming the diquark
model and compare the obtained diquark masses
those used for the hyperon production in Ref.~\cite{Andersson_PhysRept}.
We fit the production cross sections of $\Lambda_c^+$ baryons and
$\Sigma_c$ baryons using exponential functions, shown as the solid and
dashed lines in Fig.~\ref{fig:cross_mass_fit_charm}.
We obtain parameters of Eq.~\ref{eq:cross_mass} to be
$a_0=(6.2\pm 7.0)\times 10^7$~pb, 
$a_1= (-6.3\pm 0.5)/$(GeV/$c^2$) with $\chi^2/ndf=0.2/1$
for the $\Lambda_c^+$ family
and 
$a_0=(4.6\pm 12.0)\times 10^6$~pb, 
$a_1= (-5.8 \pm 1.0)/$(GeV/$c^2$) 
with $\chi^2/ndf=0.5/1$ for the $\Sigma_c$ family. 
%
%
The slope parameters for $\Lambda_c^+$ baryons and $\Sigma_c^0$ baryons
 are consistent within statistical uncertainties, and
the ratio of production cross sections of 
$\Sigma_c^0$ to $\Lambda_c^+$ baryons is 
$0.27 \pm 0.07$, using the weighted average of the slope parameters
$\langle a_1\rangle =-6.2/$(GeV/$c^2$). 
Note that the uncertainties of the $a_0$ parameters are reduced by fixing the $a_1$ parameter.
In the relativistic string fragmentation model~\cite{Andersson_PhysRept},
$q\bar{q}$ pairs are created 
in the strong color force in analogy with the Schwinger effect in QED.
Similarly, in the diquark model, a diquark and anti-diquark pair is created to form
a baryon or an anti-baryon.
Assuming that the production cross sections of charmed baryons are 
proportional to the production probability of a diquark, 
the ratio of the production cross sections of $\Lambda_c^+$ resonances and
$\Sigma_c$ resonances is proportional to 
$\exp(-\pi \mu^2/\kappa)$~\cite{Andersson_PhysicaScripta},
where $\kappa$ is the string tension, $\kappa/\pi \sim 250^2 $~(MeV$^2$),
and $\mu$ is the mass of the diquark.
The obtained mass squared difference of spin-0 and 1 diquark, 
${m}({ud}_1)^2-{m}({ud}_0)^2$,
is $(8.2 \pm 0.8) \times 10^4$~(MeV/$c^2$)$^2$. 
This is slightly higher than but consistent with the value described
in Ref.~\cite{Andersson_PhysRept},
$490^2-420^2=6.4\times 10^4$~(MeV/$c^2$)$^2$.
Our results favor the diquark model in
the production mechanism of charmed baryons and a spin-0 diquark 
component of the $\Lambda_c^+$ ground state and low-lying excited states.


%
%
%

\section{Summary}
\label{sec:Summary}

We have measured the inclusive production cross sections of hyperons and charmed 
baryons from $e^+e^-$ 
annihilation near the $\Upsilon(4S)$ energy using high-statistics data recorded at Belle.
The direct production cross section divided by the spin multiplicities for $S=-1$ hyperons 
except for $\Sigma(1385)^+$ lie on one common exponential function of mass. 
A suppression for $\Sigma(1385)^+$ and $S=-2, -3$ hyperons is observed, which is likely
due to decuplet suppression and strangeness suppression in the fragmentation.
The production cross sections of charmed baryons are significantly higher
than those of excited hyperons, and strong suppression of $\Sigma_c$
with respect to $\Lambda_c^+$ is observed.
The ratio of the production cross sections of $\Lambda_c^+$ and $\Sigma_c$ is
consistent with the difference of the production probabilities of 
spin-0 and spin-1 diquarks in the fragmentation process.
This observation supports the theory that the diquark production is 
the main process of charmed baryon production from $e^+e^-$ annihilation, 
and that the diquark structure exists in the ground state and low-lying excited
states of $\Lambda_c^+$ baryons.

\begin{acknowledgments}
We thank the KEKB group for the excellent operation of the
accelerator; the KEK cryogenics group for the efficient
operation of the solenoid; and the KEK computer group,
the National Institute of Informatics, and the
PNNL/EMSL computing group for valuable computing
and SINET5 network support.  We acknowledge support from
the Ministry of Education, Culture, Sports, Science, and
Technology (MEXT) of Japan, the Japan Society for the
Promotion of Science (JSPS), and the Tau-Lepton Physics
Research Center of Nagoya University;
the Australian Research Council;
Austrian Science Fund under Grant No.~P 26794-N20;
the National Natural Science Foundation of China under Contracts
No.~10575109, No.~10775142, No.~10875115, No.~11175187, No.~11475187,
No.~11521505 and No.~11575017;
the Chinese Academy of Science Center for Excellence in Particle
Physics;
the Ministry of Education, Youth and Sports of the Czech
Republic under Contract No.~LTT17020;
the Carl Zeiss Foundation, the Deutsche Forschungsgemeinschaft, the
Excellence Cluster Universe, and the VolkswagenStiftung;
the Department of Science and Technology of India;
the Istituto Nazionale di Fisica Nucleare of Italy;
the WCU program of the Ministry of Education, National Research
Foundation (NRF)
of Korea Grants No.~2011-0029457, No.~2012-0008143,
No.~2014R1A2A2A01005286,
No.~2014R1A2A2A01002734, No.~2015R1A2A2A01003280,
No.~2015H1A2A1033649, No.~2016R1D1A1B01010135, No.~2016K1A3A7A09005603,
No.~2016K1A3A7A09005604, No.~2016R1D1A1B02012900,
No.~2016K1A3A7A09005606, No.~NRF-2013K1A3A7A06056592;
the Brain Korea 21-Plus program, Radiation Science Research Institute,
Foreign Large-size Research Facility Application Supporting project and
the Global Science Experimental Data Hub Center of the Korea Institute
of Science and Technology Information;
the Polish Ministry of Science and Higher Education and
the National Science Center;
the Ministry of Education and Science of the Russian Federation and
the Russian Foundation for Basic Research;
the Slovenian Research Agency;
Ikerbasque, Basque Foundation for Science and
MINECO (Juan de la Cierva), Spain;
the Swiss National Science Foundation;
the Ministry of Education and the Ministry of Science and Technology of
Taiwan;
and the U.S.\ Department of Energy and the National Science Foundation.
 
\end{acknowledgments}

\appendix

\section{Reconstruction efficiency}
\label{Appendix_efficiency}

\begin{figure}[ht]
 \includegraphics[scale=0.45]{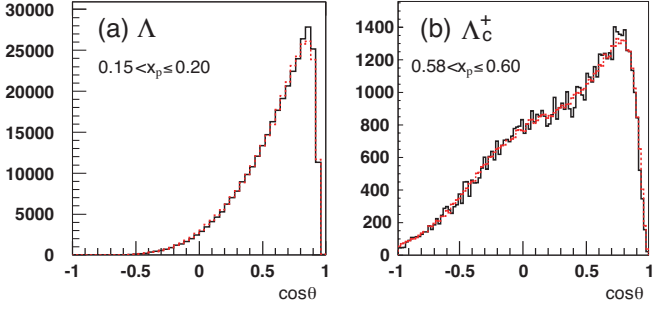}
 \caption{Polar angular distribution for $\Lambda$ (a) and $\Lambda_c^+$
 (b) in the laboratory system. Solid (dotted) histograms show the distributions of 
real (MC) data.}
\label{fig:angle_mc}
\end{figure}

\begin{figure}[ht]
 \includegraphics[scale=0.4]{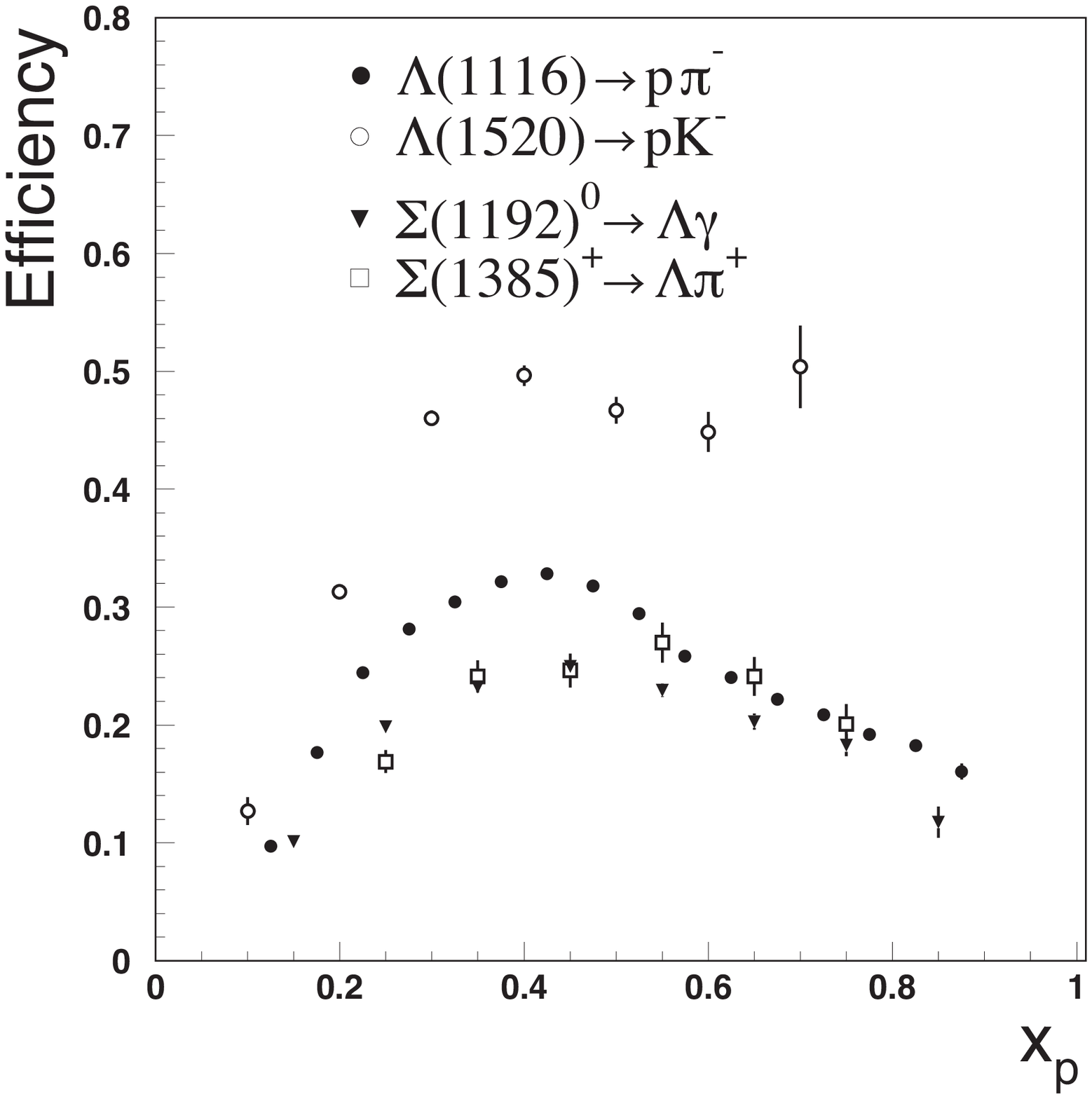}
 \caption{Reconstruction efficiencies for $S=-1$ hyperons.}
\label{fig:acc_hyp-1}
\end{figure}

\begin{figure}[ht]
 \includegraphics[scale=0.4]{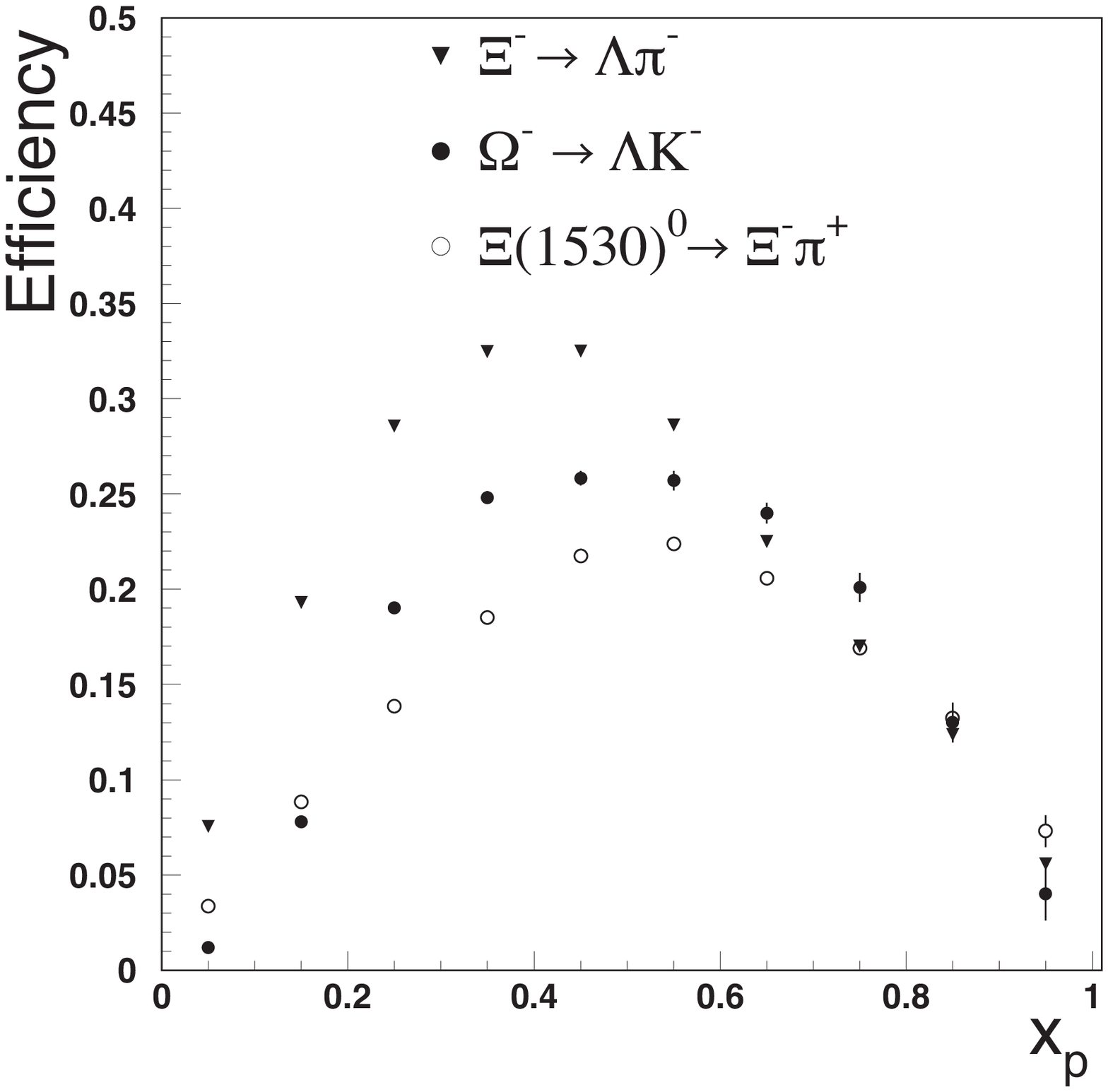}
 \caption{Reconstruction efficiencies for $S=-2,-3$ hyperons.}
\label{fig:acc_hyp-2}
\end{figure}

\begin{figure}[ht]
 \includegraphics[scale=0.4]{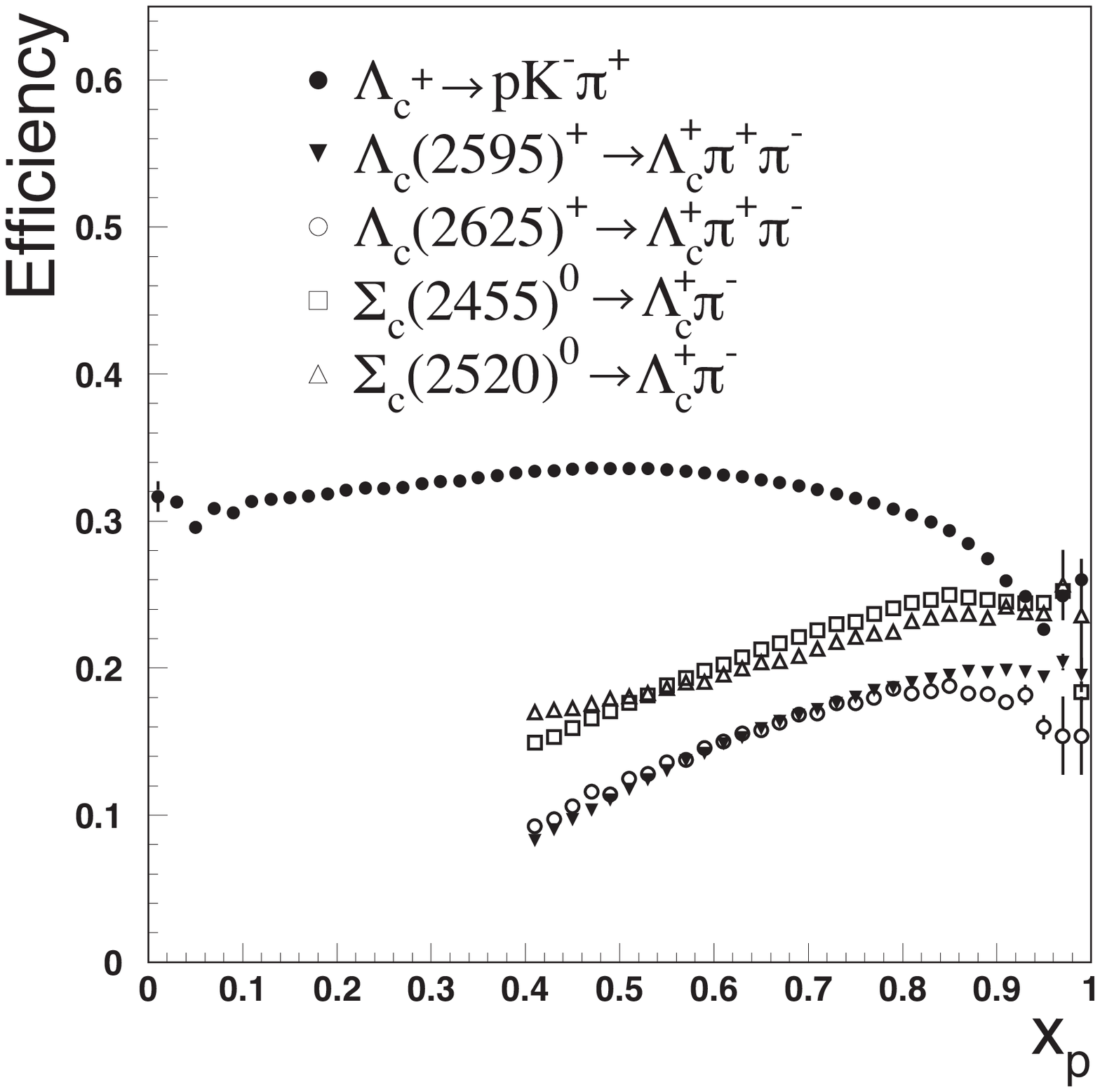}
 \caption{Reconstruction efficiencies for charmed baryons.}
\label{fig:acc_lamc}
\end{figure}

\begin{figure}[ht]
 \includegraphics[scale=0.4]{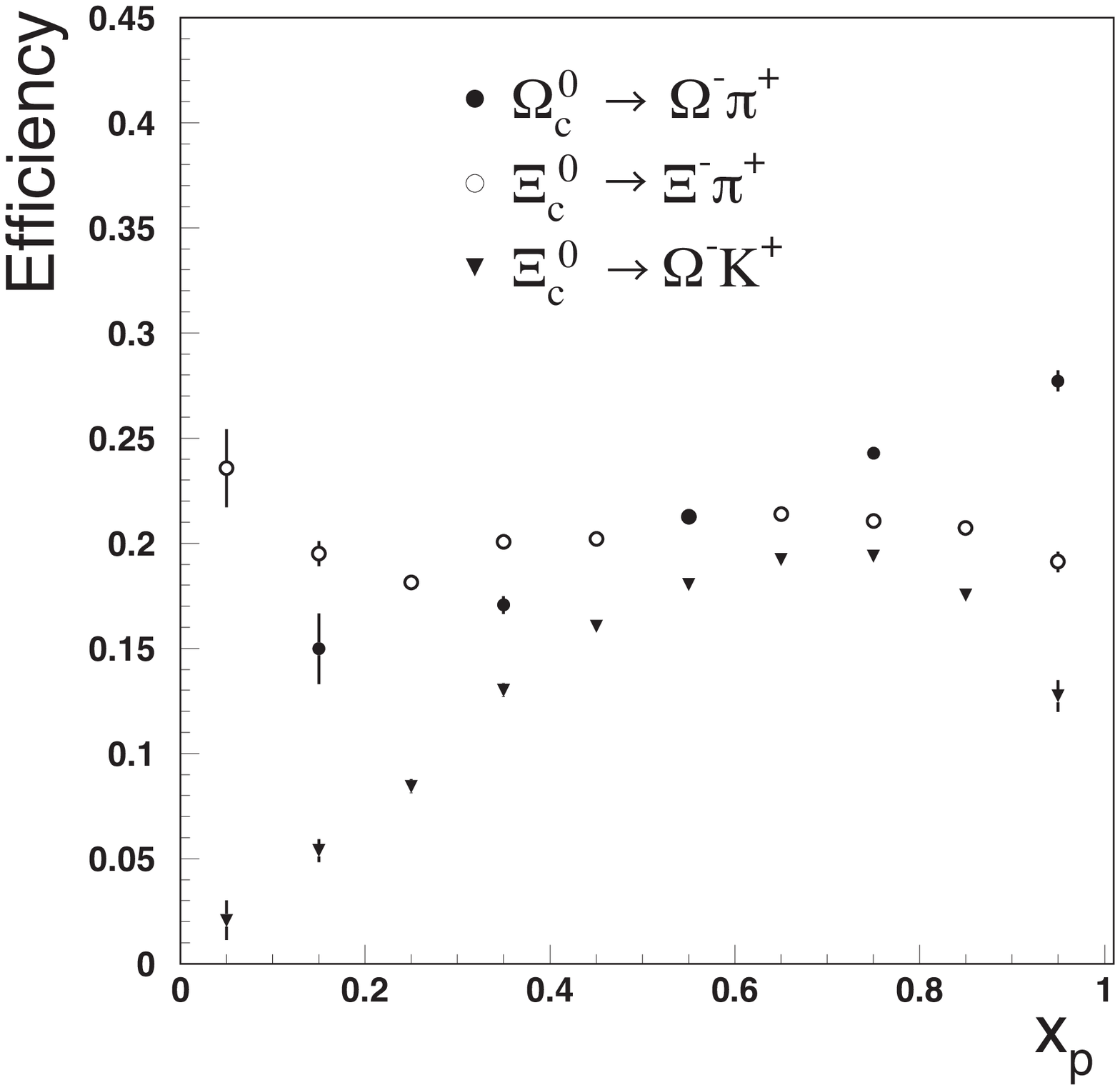}
 \caption{Reconstruction efficiencies for charmed strange baryons.}
\label{fig:acc_charm_strange}
\end{figure}

\begin{figure*}
\includegraphics[scale=0.7]{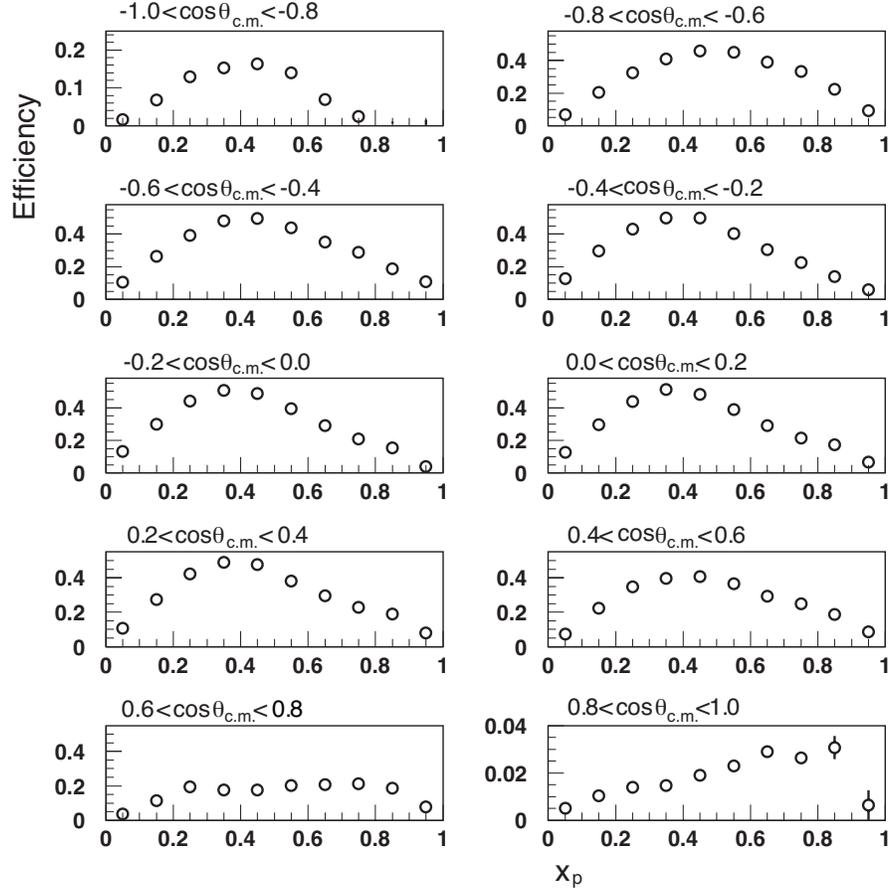}
 \caption{Reconstruction efficiencies for $\Xi^-$ hyperons in the $e^+e^-$
center of mass system.}
\label{fig:acc_xi}
\end{figure*}

\begin{figure}[ht]
 \includegraphics[scale=0.4]{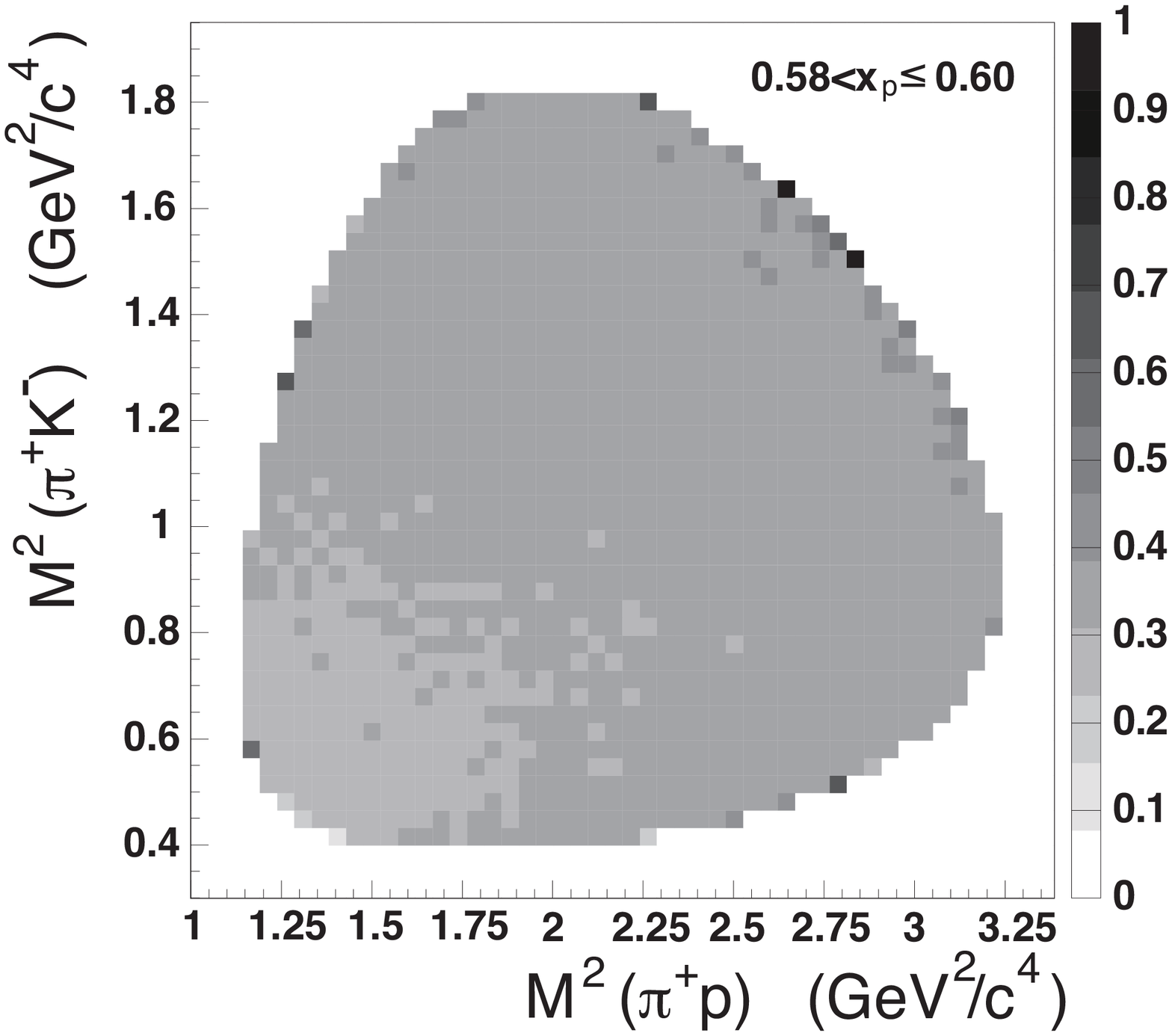}
 \caption{Reconstruction efficiency over the Dalitz plot for
 $\Lambda_c^+ \rightarrow pK^-\pi^+$.}
\label{fig:acc_lamc2286}
\end{figure}

The reconstruction efficiencies are obtained using MC event samples
that are generated using PYTHIA.
The angular distributions of each particle are well reproduced
by the MC event generator. Figure~\ref{fig:angle_mc} shows the 
polar angular distribution of the $\Lambda$ and $\Lambda_c^+$ in the 
laboratory system for the real data and MC.
The detector responses are simulated using
GEANT3 package.
In order to cancel the difference in the
momentum distribution between real and MC events, 
the corrections for the reconstruction efficiencies are applied in each
$x_p$ bin as shown in Figs.~\ref{fig:acc_hyp-1}-\ref{fig:acc_charm_strange}.

The trajectory of the $\Xi^-$ ($\Omega^-$) hyperon is reconstructed from the
momentum and vertex point of a $\Lambda \pi^-$ ($\Lambda K^-$) pair, and
the closest point with respect to the IP is obtained.
Because the reconstruction of the momentum vector of these hyperons at
the IP is complicated compared to $S=-1$ hyperons, the reconstruction efficiencies
are obtained in each angular and  $x_p$ bin.
The correction factors for $\Xi^-$ are shown in Fig.~\ref{fig:acc_xi} as
an example. 

In the $\Lambda_c^+ \rightarrow \pi^+K^-p$ decay, 
the intermediate resonances ($K(890)^0$, $\Delta$, and $\Lambda(1520)$) can
contribute as described in Sec.~\ref{sec:ana_charmed_baryons}.
To avoid the uncertainty in the reconstruction efficiency
correction due to these intermediate states, the correction is applied for the Dalitz
distribution of $\Lambda_c^+$ signal region after subtracting the
sideband events.
Fig.~\ref{fig:acc_lamc2286} shows the reconstruction efficiency over the Dalitz
plot for $\Lambda_c^+\rightarrow pK^-\pi^+$ in the region of $0.58<x_p<0.6$.

\section{feed-down from higher resonances}
\label{Appendix_feed-down}

In order to obtain the direct production cross sections, 
the feed-down contributions
from heavier states are subtracted. 
We consider all feed-down contributions that are listed in the PDG~\cite{PDG2016}.
There may be decay modes that have not yet been measured, and so 
are not listed.
Thus, the ``true'' direct cross sections may be smaller.
However, the production cross sections of heavy particles are expected to be
suppressed according to the exponential mass dependence, and feed-down
contributions from heavier particles should be small.

The feed-down contributions are summarized in
Tables~\ref{tbl:feed_lambda}--\ref{tbl:feed_sc2455}.
Table~\ref{tbl:cross_inclusive_direct} shows 
a summary of the inclusive and direct cross sections.
We use the values of the inclusive cross sections that are obtained by this
work. The branching fractions are obtained from Ref.~\cite{PDG2016}.

A preliminary measurement of the branching fraction of inclusive
$\Lambda_c^+ \rightarrow \Lambda X$ decay is found to be $0.3698 \pm 0.0218$
by BES~III~\cite{BES_Lc_Lambda}.
This inclusive branching fraction contains
$ \Lambda_c^+ \rightarrow \Sigma^0 X \rightarrow \Lambda \gamma X$ decay mode.
In order to avoid double counting of feed-down from $\Sigma^0$,
we need to eliminate the inclusive $\Lambda_c^+ \rightarrow \Sigma^0 X$
mode.
However, this decay mode has not yet been measured. If we use exclusive decay modes,
$ \Lambda_c^+ \rightarrow \Sigma^0 \pi^+ $ (1.29\%),
$ \Lambda_c^+ \rightarrow \Sigma^0 \pi^+ \pi^0 $ (2.3\%),
$ \Lambda_c^+ \rightarrow \Sigma^0 \pi^+ \pi^+ \pi^- $ (1.13\%),
 $\Lambda_c^+ \rightarrow \Lambda X$ becomes 
32.26\%.
The amount of feed-down from $\Lambda_c^+$ to $\Lambda$ is estimated as
$141.79\times 0.3226=45.74$~pb.
The sum of the feed-down from $\Lambda_c^+$ listed in
Table~\ref{tbl:feed_lambda} is 32.17~pb.
We take the difference of these two values, $45.74-32.17=13.57$~pb, 
as the systematic uncertainty for the feed-down from $\Lambda_c^+$ to $\Lambda$.


The branching fraction of 
${\cal B}(\Lambda_c(2595)^+ \rightarrow \Lambda_c^+\pi^+\pi^-)$ is obtained to be
$0.346 \pm 0.012 \ ({\rm syst.})$ using Cho's function~\cite{ChoPRD50}
with the parameter obtained by CDF~\cite{CDF_PRD84012003}.
In this calculation, we integrate the mass spectrum of
$\Lambda_c(2595)^+$ in the range of $0.28$~GeV/$c^2$$ < \Delta M(\pi\pi) <
0.33$~GeV/$c^2$, and estimate the uncertainty by changing the mass range with
$\pm 5$~MeV, which is conservatively larger than the mass resolution.
Taking into account the world-average relative branching fraction of
  ${\cal B}(\Sigma_c(2455)^0\pi^+) /
  ({\cal B}(\Sigma_c(2455)^0\pi^+)+{\cal B}(\Sigma_c(2455)^{++}\pi^-)+{\cal B}(${non-resonant}$\,
  \Lambda_c^+\pi^+\pi^-))
  = 0.36 \pm 0.10$, 
we obtain ${\cal B}(\Lambda_c(2595)^+\rightarrow \Sigma_c(2455)^0\pi^+)=0.125\pm 0.034$.

\begin{table}[htb]
  \caption{Feed-down to $\Lambda$. 
For the sum of the systematic uncertainties of
the feed-down from $\Lambda_c^+$, the difference of 
 the branching fractions of inclusive $\Lambda_c^+\rightarrow \Lambda X$ and
exclusive decay modes is used as described in the text.}
  \label{tbl:feed_lambda}
\begin{tabular}{ l c r } \hline\hline
Decay mode & Branching & Feed-down (pb) \\
           & fraction  &  \\
\hline
$\Sigma(1192)^0\rightarrow\Lambda\gamma $  & 1 & $ 63.44\pm0.66\pm3.2$ \\
$\Sigma(1385)^{\pm,0}\rightarrow\Lambda\pi^{\pm,0} $  & $0.87\pm0.015$ & $ 57.99\pm0.92\pm6.5$ \\
$\Sigma(1385)^0\rightarrow\Lambda\gamma $  & $0.0125\pm0.0012$ & $ 0.278\pm0.004\pm0.041$ \\
$\Xi^{-,0}\rightarrow\Lambda\pi^{-,0} $  & $0.99887\pm0.00035$ & $ 32.35\pm0.32\pm1.5$ \\
$\Lambda(1520)\rightarrow\Lambda\pi\pi $  & $0.0222\pm0.0091$ & $0.284\pm0.005\pm0.12$ \\
$\Lambda(1520)\rightarrow\Lambda\gamma $  & $0.0085\pm0.0015$ & $0.109\pm0.002\pm0.021$ \\
$\Omega^-\rightarrow\Lambda\,K^- $  & $0.678\pm0.007$ & $ 0.601\pm0.011\pm0.028$ \\
$\Lambda_c^+\rightarrow\Lambda\pi^+ $  & $0.013\pm0.0007$ & $ 1.76\pm0.01\pm0.14$ \\
$\Lambda_c^+\rightarrow\Lambda\pi^+\pi^0 $  & $0.071\pm0.0042$ & $ 10.07\pm0.058\pm0.77$ \\
$\Lambda_c^+\rightarrow\Lambda\pi^+\pi^-\pi^+ $  & $0.0381\pm0.003$ & $ 5.402\pm0.031\pm0.50$ \\
$\Lambda_c^+\rightarrow\Lambda\pi^+\pi^-\pi^+\pi^0 $  & $0.023\pm0.008$ & $ 3.261\pm0.019\pm1.2$ \\
$\Lambda_c^+\rightarrow\Lambda\pi^+\eta $  & $0.024\pm0.005$ & $ 3.40\pm0.02\pm0.73$ \\
$\Lambda_c^+\rightarrow\Lambda\pi^+\omega $  & $0.016\pm0.006$ & $ 2.269\pm0.013\pm0.86$ \\
$\Lambda_c^+\rightarrow\Lambda\,K^+\overline{K^0} $  & $0.0057\pm0.0011$ & $ 0.808\pm0.005\pm0.16$ \\
$\Lambda_c^+\rightarrow\Lambda\,K^+ $  & $(7\pm 1)\times 10^{-4}$ & $ 0.098\pm0.001\pm0.02$ \\
$\Lambda_c^+\rightarrow\Lambda\,e^+\nu_e $  & $0.036\pm0.004$ & $ 5.104\pm0.029\pm0.62$ \\
Sum & & $185.3\pm2.2\pm16$\\
\hline\hline
\end{tabular}
\end{table}

\begin{table}[htb]
  \caption{Feed-down to $\Sigma^0$.}
  \label{tbl:feed_sigma0}
\begin{tabular}{ l c r } \hline\hline
Decay mode & Branching & Feed-down (pb) \\
           & fraction  &  \\
\hline
$\Sigma(1385)^{\pm}\rightarrow\Sigma^0\pi^{\pm} $  & $0.117\pm0.015$ &$ 2.600\pm0.041\pm0.44$ \\
$\Lambda(1520)\rightarrow\Sigma^0\pi^0 $  & $0.14\pm0.0033$ & $1.792\pm0.029\pm0.13$ \\
$\Lambda_c^+\rightarrow\Sigma^0\pi^+ $  & $0.0127\pm0.0009$ & $ 1.801\pm0.010\pm0.15$ \\
$\Lambda_c^+\rightarrow\Sigma^0\pi^+\pi^0 $  & $0.025\pm0.009$ & $ 3.545\pm0.020\pm1.3$ \\
$\Lambda_c^+\rightarrow\Sigma^0\pi^+\pi^+\pi^- $  & $0.0113\pm0.0001$ & $ 1.602\pm0.009\pm0.08$ \\
$\Lambda_c^+\rightarrow\Sigma^0K^+ $  & $0.0006\pm0.0001$ & $ 0.08\pm0.0005\pm0.02$ \\
Sum & & $11.157\pm0.055\pm1.433$\\
\hline\hline
\end{tabular}
\end{table}

\begin{table}[htb]
  \caption{Feed-down to $\Sigma(1385)^+$.}
  \label{tbl:feed_sigma1385}
\begin{tabular}{ l c r } \hline\hline
Decay mode & Branching & Feed-down (pb) \\
           & fraction  &  \\
\hline
$\Lambda(1520)\rightarrow\Sigma(1385)^+\pi^- $  & $0.0137\pm0.0017$ &$ 0.175\pm0.003\pm0.025$ \\
$\Lambda_c^+\rightarrow\Sigma(1385)^+\eta $  & $0.0108\pm0.0032$ & $ 1.531\pm0.007\pm0.46$ \\
$\Lambda_c^+\rightarrow\Sigma(1385)^+\pi^+\pi^- $  & $0.01\pm0.005$ & $ 1.418\pm0.003\pm0.71$ \\
$\Lambda_c^+\rightarrow\Sigma(1385)^+\rho^0 $  & $0.005\pm0.004$ & $ 0.709\pm0.006\pm0.57$ \\
Sum & & $3.833\pm0.013\pm1.3$\\
\hline\hline
\end{tabular}
\end{table}

\begin{table}[htb]
  \caption{Feed-down to $\Lambda(1520)$, $\Xi^-$, and $\Xi(1530)^0$.}
  \label{tbl:feed_l1520_xi}
\begin{tabular}{ l c r } \hline\hline
Decay mode & Branching & Feed-down (pb) \\
           & fraction  &  \\
\hline
$\Lambda_c^+\rightarrow\Lambda(1520)\pi^+ $  & $0.024\pm0.006$ & $ 3.40\pm0.02\pm0.87$ \\
Sum & & $3.40\pm0.02\pm0.87$\\
$\Xi(1530)^{0,-}\rightarrow\Xi^-\pi^{+,0} $  & 0.5 & $ 3.855\pm0.062\pm0.18$ \\
$\Omega^-\rightarrow\Xi^-\pi^{0} $  & $0.086\pm0.004$ & $ 0.076\pm0.001\pm0.004$ \\
$\Lambda_c^+\rightarrow\Xi^-K^+\pi^+ $  & $0.007\pm0.0008$ & $ 0.993\pm0.006\pm0.049$ \\
Sum & & $4.924\pm0.063\pm0.23$\\
$\Lambda_c^+\rightarrow\Xi(1530)^0K^+ $  & $0.0033\pm0.0009$ & $ 0.936\pm0.005\pm0.046$ \\
Sum & & $0.936\pm0.005\pm0.046$\\
\hline\hline
\end{tabular}
\end{table}

\begin{table}[htb]
  \caption{Feed-down to $\Lambda_c^+$.}
  \label{tbl:feed_lc}
\begin{tabular}{ l c r } \hline\hline
Decay mode & Branching & Feed-down (pb) \\
           & fraction  &  \\
\hline
$\Lambda_c(2595)^+\rightarrow\Lambda_c^+\pi\pi $  & 1 & $ 10.157\pm0.011\pm0.88$ \\
$\Lambda_c(2625)^+\rightarrow\Lambda_c^+\pi\pi $  & 1 & $ 15.37\pm0.12\pm1.3$ \\
$\Sigma_c(2455)^{0,+,++}\rightarrow\Lambda_c^+\pi^{-,0,+} $  & 1 & $ 20.09\pm0.21\pm2.8$ \\
$\Sigma_c(2520)^{0,+,++}\rightarrow\Lambda_c^+\pi^{-,0,+} $  & 1 & $ 23.30\pm0.34\pm3.2$ \\
$\Sigma_c(2800)^{0,+,++}\rightarrow\Lambda_c^+\pi^{-,0,+} $  & 1 & $ 5.3\pm1.1\pm3.2$ \\
Sum & & $74.195\pm1.206\pm5.571$\\
\hline\hline
\end{tabular}
\end{table}

\begin{table}[htb]
  \caption{Feed-down to $\Sigma_c(2455)^0$.}
  \label{tbl:feed_sc2455}
\begin{tabular}{ l c r } \hline\hline
Decay mode & Branching & Feed-down (pb) \\
           & fraction  &  \\
\hline
$\Lambda_c(2595)^+\rightarrow\Sigma_c(2455)^0\pi^+ $  & $0.125\pm0.035$ & $ 1.266\pm0.001\pm0.37$ \\
Sum & & $1.266\pm0.001\pm0.371$\\
\hline\hline
\end{tabular}
\end{table}

\clearpage


\end{document}